\documentclass[10pt,aps,floatfix,nofootinbib]{revtex4}

\usepackage{graphicx}
\usepackage{amssymb}
\usepackage{amsmath}
\usepackage{amsfonts}
\usepackage{epstopdf}
\usepackage{epsfig}
\usepackage{wrapfig}

\newtheorem{theorem}{Theorem}

\newtheorem{definition}[theorem]{Definition}

\newtheorem{lemma}[theorem]{Lemma}

\newenvironment{proof}[1][Proof]{\noindent\textbf{#1.} }{\ \rule{0.5em}{0.5em}}

\newtheorem{myproposition}{Proposition}
\newtheorem{mylemma}{Lemma}
\newtheorem{mydefinition}{Definition}
\newtheorem{mytheorem}{Theorem}

\newtheorem{mycorollary}{Corollary}

\begin{document}

\title{On the Exact Evaluation of Certain Instances of the Potts Partition
Function by Quantum Computers}
\author{Joseph Geraci}
\affiliation{Department of Mathematics, University of Toronto, Toronto, ON M5S 2E4, Canada}
\affiliation{Chemistry Department, University of Southern California, CA 90089, USA}
\author{Daniel A. Lidar}
\affiliation{Departments of Chemistry, Electrical Engineering, and Physics, University of
Southern California, CA 90089, USA}

\begin{abstract}
We present an efficient quantum algorithm for the exact evaluation of either the fully
ferromagnetic or anti-ferromagnetic $q$-state Potts partition function
$Z$ for a
family of graphs related to irreducible cyclic codes. This problem is related to the evaluation of the Jones and
Tutte polynomials. We consider the connection between the weight enumerator
polynomial from coding theory and $Z$ and exploit the fact that there exists
a quantum algorithm for efficiently estimating Gauss sums in order to obtain
the weight enumerator for a certain class of linear codes. In this way we
demonstrate that for a certain class of sparse graphs, which we call Irreducible
Cyclic Cocycle Code (ICCC$_{\epsilon }$) graphs, quantum computers 
provide a polynomial speed up in the difference between the number of
edges and vertices of the graph, and an exponential speed up in $q$, over the
best classical algorithms known to date.
\end{abstract}

\maketitle

\section{Introduction}

A wealth of results has been obtained since the dramatic early results \cite%
{Shor:97,Grover:96} on quantum speedups relative to classical algorithms. A
relatively unexplored field is quantum algorithms for problems in classical
statistical mechanics. The earliest contribution to this subject \cite%
{Lidar:PRE97a} obtained a modest speedup in that it avoided critical slowing
down \cite{Swendsen87} in the problem of sampling from the Gibbs
distribution for Ising spin glass models. Subsequently Ref. \cite{Lidar:QWGT}
raised the question of providing a classification of classical statistical
physics problems in terms of their quantum computational complexity. In this
work we shed light on this classification by considering the problem of
evaluating the Potts model partition function $Z$ for classical spin systems
on graphs. It is known that under particular conditions even certain
approximations for $Z$ are unlikely to be efficient, barring an $NP=RP$
surprise \cite{Welsh}. Here we present a class of sparse graphs (which we call $%
\mathrm{ICCC}_{\epsilon }$) for which \emph{exact} quantum evaluation of $Z$
is possible with a polynomial speedup over the best classical algorithms
available to date. 

\subsection{The Potts Model}

\label{Potts}

Let $\Gamma =(E,V)$ be a weighted graph with edge set $E$ and vertex set $V$%
. The $q$-state Potts model is a generalization of the Ising model where a $q
$-state spin resides on each vertex. In the Ising model $q=2$, whereas in
the Potts model $q\geq 2$. The edge connecting vertices $i$ and $j$ has
weight $J_{ij}$, which is also the interaction strength between the
corresponding spins. The Potts model Hamiltonian for a particular spin
configuation $\sigma =(\sigma _{1},...,\sigma _{|V|})$ is

\begin{equation}
H(\sigma )=-\sum\limits_{<ij>}J_{ij}\delta _{\sigma _{i}\sigma _{j}}
\label{eq:H}
\end{equation}%
where summation is over nearest neighbors, and where $\delta _{\sigma
_{i}\sigma _{j}}=1$ ($0$) if $\sigma _{i}=\sigma _{j}$ ($\sigma _{i}\neq
\sigma _{j}$). Thus only nearest neighbor parallel spins contribute to the
energy. The probability $P(\sigma )$, of finding the spin in the Potts model
in some configuration $\sigma $ at a given temperature $T$, is given by the
Gibbs distribution

\begin{equation}
P(\sigma )=\frac{e^{-\beta H(\sigma )}}{Z(\beta )},
\end{equation}%
where $\beta =1/(k_{\mathrm{B}}T)$ is the inverse temperature in energy
units, and $k_{\mathrm{B}}$ is the Boltzmann constant. The normalization
factor is the \emph{partition function}

\begin{equation}
Z(\beta )=\sum\limits_{\{\sigma \}}e^{-\beta H(\sigma )},  \label{partition}
\end{equation}%
which plays a central role in statistical physics, since many thermodynamic
quantities can be derived from it \cite{Reichl}. When for all configurations 
$\beta \ll |H(\sigma )|$, the probability distribution becomes flat: $%
P(\sigma )\approx 1/Z(\beta )$, so that at high temperatures randomness
dominates.

The partition function can be rewritten as a polynomial:

\begin{eqnarray}
Z(\beta ) &=&\sum\limits_{\{\sigma \}}e^{\beta \sum_{<ij>}J_{ij}\delta
_{\sigma _{i}\sigma _{j}}}=\sum\limits_{\{\sigma
\}}\prod\limits_{<ij>}e^{\beta J_{ij}\delta _{\sigma _{i}\sigma _{j}}} 
\notag \\
&=&\sum\limits_{\{\sigma \}}\prod\limits_{<ij>}(1+v_{ij}(\beta )\delta
_{\sigma _{i}\sigma _{j}}),  \label{pottspart}
\end{eqnarray}%
where%
\begin{equation}
v_{ij}(\beta )=e^{\beta J_{ij}}-1.  \label{eq:v}
\end{equation}

Now let us consider the case when the interactions $J_{ij}$ are a constant $%
J $. Then the Hamiltonian (\ref{eq:H}) of this system can be written as

\begin{equation}
H(\sigma )=-J|U(\sigma )|
\end{equation}%
where $U(\sigma )$ is the subset of edges whose vertices have the same spin
for a particular spin configuration $\sigma $, and $|U(\sigma )|$ is the
number of such subsets. If we let 
\begin{equation}
y=e^{-\frac{J}{kT}}
\end{equation}%
we can write the Potts partition function as

\begin{equation}
Z(y)=\sum_{\sigma }y^{-|U(\sigma )|}.  \label{newpotts}
\end{equation}

\subsection{Relation between the Potts partition function, knot invariants,
and graph theory}

There is a rich inter-relation between classical statistical mechanics and
topology, in particular, the theory of the classification of knots. The
first such connection was established by Jones \cite{Jones:89}, who
discovered the second knot invariant (the Jones polynomial (a Laurent
polynomial), after the Alexander polynomial) during his investigation of the
topological properties of braids \cite{Jones:85}. It is known that the
classical evaluation of the Jones polynomial is $\sharp $P-hard \cite%
{Jaeger:90}.

The connection between knots and models of classical statistical mechanics
was embellished by Kauffman \cite{Kauffman:book}. Knot invariants are, in
turn, also tightly related to graph theory; e.g., the graph coloring problem
can be considered an instance of evaluation of the Kauffman bracket
polynomial, via the Tutte polynomial \cite{Kauffman:book,Alon:95}. The $q$%
-state Potts partition function on a graph $\Gamma $ is connected to the
Tutte polynomial $\mathcal{T}_{\Gamma }$ for the same graph via 
\begin{equation}
Z_{\Gamma }(v)=q^{n}\mathcal{T}_{\Gamma }(\frac{q+v}{v},v+1)  \label{tutte}
\end{equation}%
where as in Eq. (\ref{eq:v}), $v+1=e^{-\beta }$. This means that the Potts
partition function is equivalent to some easily computed function times the
Tutte polynomial along the hyperbola $\mathcal{H}_{q}=(x-1)(y-1)=q$. But for
planar graphs, when $q>2$ the Tutte polynomial is $\sharp $P-hard to
evaluate at points along $\mathcal{H}_{q}$ \cite{Welsh}. For a review of the
connection between the Potts partition function and the various polynomials
mentioned above, see \cite{Kauffman:book} and also \cite%
{Nechaev:98,Lidar:QWGT}. It immediately follows from Eq.~(\ref{tutte}) and
complexity results concerning the Tutte polynomial, that the evaluation of
the Potts partition function is also $\sharp $P-hard. It is not known
whether there is an fpras (fully polynomial randomized approximation scheme) 
\cite{Welsh} for the $q$-state fully ferromagnetic Potts partition function,
but it is known that if there is an fpras for the fully anti-ferromagnetic
Potts partition function then $NP=RP$ \cite{Welsh} and therefore it seems
unlikely that an fpras will be found for this case.

\subsection{Previous quantum complexity results}

The first connection between knots and quantum mechanics was established by
Witten, who showed that the Jones polynomial can be expressed in terms of a
topological quantum field theory \cite{Witten:88}. Recently this connection
was extended to the possibility of efficient evaluation of the Jones
polynomial by Freedman and co-workers, after showing that quantum computers
can efficiently simulate topological quantum field theory \cite{Freedman:00}%
. More specifically, there are recent results demonstrating the efficacy of
quantum computers in approximating the Jones polynomial at primitive roots
of unity \cite{Freedman:01,Aharonov:06,Wocjan:06}. In Ref. \cite{Freedman:01}
tools from topological quantum field theory \cite{Witten:88} were utilized
and it was shown that approximating the Jones polynomial at primitive roots
of unity is BQP-complete, but no explicit algorithm was provided. More
recently in \cite{Aharonov:06}, a combinatorial approach was taken which
yielded an explicit quantum algorithm and which extended the results in \cite%
{Freedman:01} to all primitive roots of unity. This leads one to hypothesize
that quantum computers will also be efficient at estimating partition
functions. Indeed, an immediate corollary of the results in \cite%
{Freedman:01,Aharonov:06,Wocjan:06}, is that the Potts partition function
over any planar graph can be approximated efficiently on a quantum computer
at certain \emph{imaginary} temperatures (see also \cite{Lidar:QWGT}). This
follows by noting that in order to obtain an equality between the Potts
partition function and the Jones polynomial (up to multiplication by an
easily computed function), the Jones variable $t$ and the temperature $T$
must be related by $t=-e^{\pm J_{\pm }/k_{\mathrm{B}}T}$ \cite{Kauffman:book}. With $t$
a root of unity ($t=e^{\frac{2\pi i}{r}}$) we then find: 
\begin{equation*}
T=i\frac{J_{\pm }r}{k_{\mathrm{B}}\pi (2+r)},\quad r\in \mathbb{N}.
\end{equation*}%
This result is of interest mainly in light of quantum Monte Carlo
simulations \cite{monte}, where one retrieves real time \emph{dynamics} from
a simulation in terms of imaginary time, via analytic continuation. Perhaps
a similar extrapolation can be achieved here between imaginary and real
temperature dynamics. While this is interesting, here we are concerned with 
\emph{thermodynamics}, and hence evaluations of the Potts partition function
at physically relevant, real temperatures.

Most closely related to our work is the very recent result due to Aharonov 
\textit{et al}. \cite{Aharonov:07} who -- generalizing Temperley-Lieb
algebra representations used in \cite{Aharonov:06} -- provided a quantum
algorithm for the additive approximation of the Potts partition function
(and other points of the Tutte plane) for any planar graph with any set of
weights on the edges. These results are the most impressive to date in
the context of approximate evaluations of the Potts partition
function, but are also subject to certain caveats. To quote
from the abstract of Ref. \cite{Aharonov:07}:
``Additive approximations are tricky; the range of the possible outcomes, might be smaller than the
size of the approximation window, in which case the outcome is meaningless. Unfortunately, ruling
out this possibility is difficult: If we want to argue that our algorithms are meaningful, we have to
provide an estimate of the scale of the problem, which is difficult here exactly because no efficient
algorithm for the problem exists!''. And: ``The case of the Potts model parameters deserves special attention. Unfortunately, despite being
able to handle non-unitary representations, our methods of proving universality seem to be non-
applicable for the physical Potts model parameters. We can provide only weak evidence that our
algorithms are non-trivial in this case, by analyzing their performance for instances for which classical
efficient algorithms exist. The characterization of the quality of the algorithm for the Potts parameters
is thus left as an important open problem.'' Finally, quoting from
Section 1.5 of Ref.~\cite{Aharonov:07}: ``Proving anything about the complexity of our algorithm for the Potts model, remains a very important
open problem. It is still possible that this case of the Tutte polynomial, with our additive approximation
window, can be solved by an efficient classical algorithm.'' To
summarize, Ref.~\cite{Aharonov:07} leaves as an open problem the
complexity of physical instances (real temperature, positive partition
function) under the restriction of an additive approximation. Nor is it
clear whether the algorithm found in Ref.~\cite{Aharonov:07} provides a
quantum speedup. The authors state: ``We believe that the main achievement here is that we demonstrate how to handle non-unitary
representations, and in particular, we are able to prove universality using non-unitary matrices.''

Recently Ref. \cite{Nest:06} gave a scheme for studying the partition
function of classical spin systems including the Potts and Ising model.
Their approach involves transforming the problem of evaluating the partition
function into the evaluation of a probability amplitude of a quantum
mechanical system and then using classical techniques to extract the
pertinent information. In essence their method involves moving into a
quantum mechanical formalism to obtain a classical result. The scheme is
therefore classical and not a quantum algorithm. 

In addition, two purely classical results should be mentioned here. One is a
state of the art result by Hartmann \cite{Hartmann:05}, who provides an
algorithm which is well suited to large ferromagnetic systems for either the
Potts or Ising model. We do not know the exact complexity of this algorithm,
however. The approach taken in our work is to utilize the connection between
classical coding theory and the partition function. For this reason we
mention the classical algorithm given in \cite{Denef:04} for calculating the
Zeta function of certain curves. This is also a state of the art algorithm
and it can be used to find the Potts partition function via the scheme we
present in this paper, though it is slower than using quantum resources. 

A quantum algorithm for finding the Zeta function of a curve is given in 
\cite{Kedlaya:05}. One could replace the role that the Gauss sum estimation 
\cite{van} plays in our scheme with this quantum algorithm for the Zeta
function. It seems that using Gauss sums is more efficient but further work
is required to make this conclusive.

\section{A Theorem about Quantum Computation and certain instances of the
Potts Model}

\subsection{Main Theorem}

We present here a polynomial time quantum algorithm for the exact evluation
of the $q$-state (fully ferromagnetic or anti-ferromagnetic) Potts partition
function $Z$ for a certain class of graphs. This class of graphs, which we
call \textquotedblleft Irreducible Cyclic Cocycle Code\textquotedblright\ $%
\mathrm{ICCC}_{\epsilon }$ graphs, comprises graphs whose incidence matrices
generate certain cyclic codes. This and other concepts used below are given
precise definitions in Section \ref{back}. The key ingredients used are the
connection of $Z$ to the weight enumerators of codes \cite{barg} and a
quantum algorithm for the approximation of Gauss sums \cite{van}.

The overall structure of the algorithm is the following:

\begin{enumerate}
\item Given a graph $\Gamma =(E,V)$, first determine if $\Gamma $ belongs to
the $\mathrm{ICCC}_{\epsilon }$ class. This decision problem can be solved
efficiently using the quantum discrete log algorithm \cite{Shor:97}. If $%
\Gamma \in \mathrm{ICCC}_{\epsilon }$ proceed to step 2, else the algorithm
may not evaluate $Z_{\Gamma }$ efficiently.

\item Identify the linear code $C(\Gamma )$ for which we shall determine the
weight spectrum.

\item Using the quantum Gauss sum estimation algorithm find the weight
spectrum of the words in $C$. This step is believed to be classically hard
but the exact complexity is unknown. It is known, however, that this step is
at least as hard as determining discrete log \cite{van}. This is the most
expensive step of the algorithm due to the large number of words one has to
deal with. This is because the number of possible spin configurations grows
exponentially in the number of vertices.

\item Take a tally of the weight spectrum obtained in the previous step.
Grover's search algorithm can be used to give an additional quadratic speed
up but this does not help in reducing the overall complexity since the
computational cost of step 3 is greater than that of the current step.

\item Using the relation given by equation (\ref{pottsA}) between the weight
spectrum of a code and $Z$, use the tally from the previous step to obtain $%
Z $ (for graphs in $\mathrm{ICCC}_{\epsilon }$).
\end{enumerate}

We now give the main theorem, after the definition of the family of graphs
for which the scheme applies.

\begin{definition}
\label{iccc}\textbf{($\mathrm{ICCC}_{\epsilon }$)} Given a constant $%
\epsilon <1$, $\mathrm{ICCC}_{\epsilon }$ is the family of graphs whose
cycle matroid matrix (CMM) representation generates a cyclic code whose dual
is irreducible cyclic of dimension $k$ and length $n$, such that 
\begin{equation}
  n=\frac{q^{k}-1}{\alpha k^{s(k)}}
  \label{eq:n}
\end{equation}%
(where $\alpha \in \mathbf{R}$ is chosen so that $n\in \mathbf{N}$ and where 
$s(k)$ is an arbitrary function whose role will be clarified below) and 
\begin{equation}
\theta _{n,k}=\frac{1}{q-1}\min_{0<j\leq \alpha k^{s(k)}}S^{\prime }(jn)
\label{eq:theta}
\end{equation}%
(where $S^{\prime }(x)$ is the sum of the digits of $x$ in base $q$) so that 
\begin{equation}
\epsilon \leq \frac{q^{\theta _{n,k}-1}}{4\sqrt{q^{k}}}.
\end{equation}
\end{definition}

Below we define the concepts entering this definition and clarify the role
of $\theta _{n,k}$ and of the bound on $\epsilon $.

We work in units such that the Boltzmann constant $k_{\mathrm{B}}=1$.

\begin{theorem}
\label{th 1}\textbf{(Main Theorem)} Let $\Gamma =(E,V)$ be a graph, $n=|E|$
and $k=|E|-|V|+c(\Gamma )$ where $c(\Gamma )$ is the number of connected
components of $\Gamma $. A quantum computer can return the exact $q$-state
fully anti-ferromagnetic or ferromagnetic Potts partition function $%
Z_{\Gamma }$ for graphs in $\mathrm{ICCC}_{\epsilon }$. For each family ($%
\epsilon $ fixed), the overall running time is $O(\frac{1}{\epsilon }%
k^{2\max [1,s(k)]}(\log q)^{2})$ and the success probability is at least $%
1-\delta $, where $\delta =[2((q^{k}-1)^{2}\epsilon -2]^{-1}$.
\end{theorem}

\begin{figure}
\epsfxsize=14cm
\centerline{\epsffile{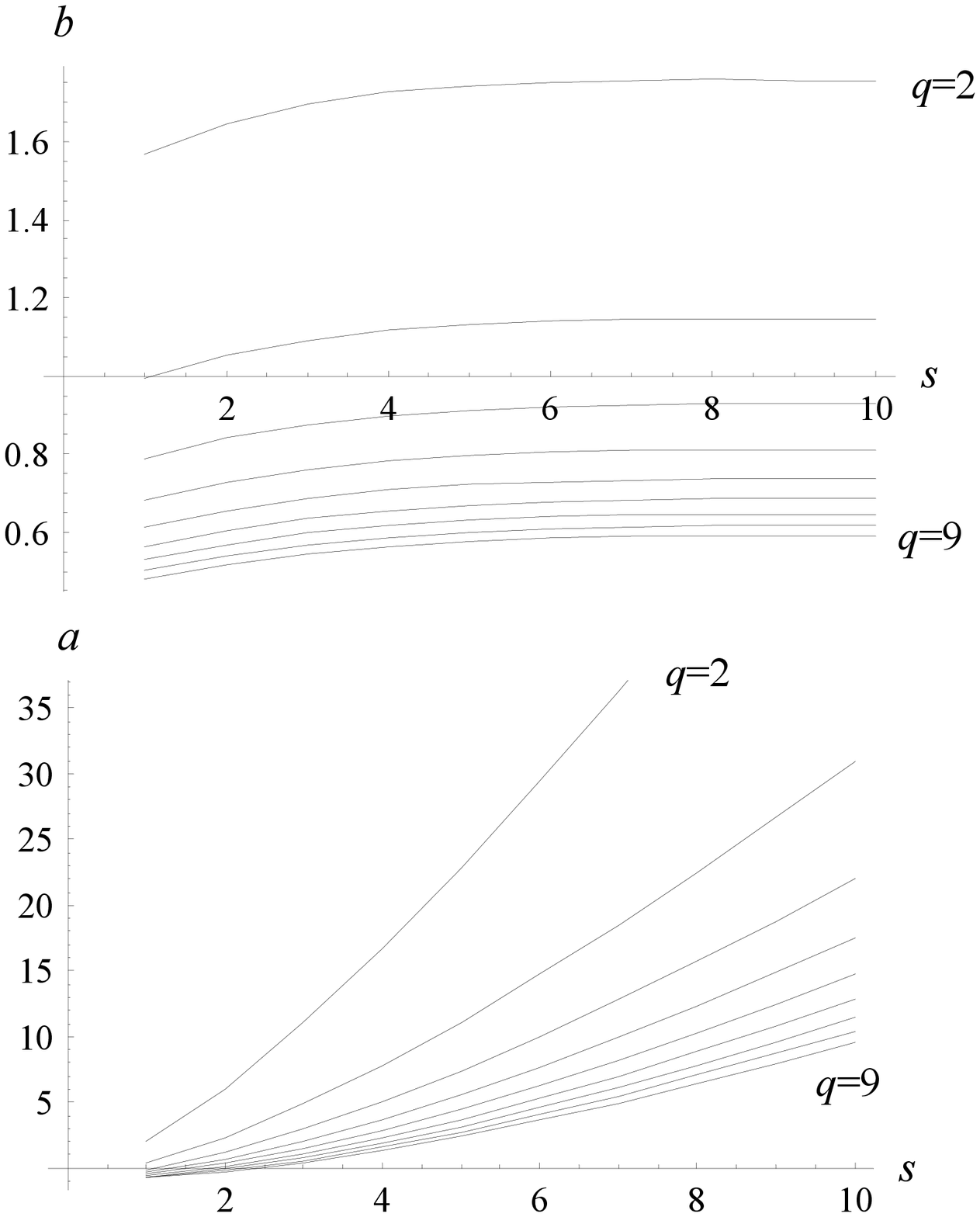}}
\caption{Coefficients $a$ and $b$ as a function of $s$, for different
  values of $q$. Here $c(\Gamma)=1$. See text for details.}
\label{fig:ab}
\end{figure}

Some remarks:

\begin{enumerate}
\item The function $s(k)$ determines the complexity of the schemes. If $%
s(k)=c\in \mathbf{R}$ (constant) then we have a polynomial time algorithm
for the exact evaluation of $Z$ for each family $\mathrm{ICCC}_{\epsilon }$.
This restriction is reflected in the graphs by enforcing that
$n=O(q^k/k^s)$, i.e., that the number of edges ($n$) and vertices
($n-k$) is close. We have numerically solved for the number of edges
$|E|$ as a function of the number of vertices $|V|$, given by the corresponding transcendental equation $|E|=|V|-c(\Gamma)+\log_q
(|E|(|E|-|V|+c(\Gamma))^s+1)$ [Eq.~(\ref{eq:n})]. A numerical fit
reveals that to an excellent approximation
\begin{equation}
  |E|=|V|+a+b \log |V|,
  \label{eq:EV}
\end{equation}
where the constants $a$ and $b$ depend on $q$ and $s$, and both increase
slowly with $s$, and decrease with $q$, as shown in
Fig.~\ref{fig:ab}. By direct substitution of Eq.~(\ref{eq:EV})
into the above transcendental equation it can be seen that the
analytical solution will have a a correction of order
$\log\log(|V|)$ to the right-hand side of
Eq.~(\ref{eq:EV}). The fact that there are
logarithmically more edges that vertices in the graphs that are
members of $\mathrm{ICCC}_{\epsilon }$ is the reason we call these
graphs sparse. The important point is that there are families of graphs for which there exist exact
polynomial-time evaluation schemes via the methods presented in this paper.
As we show below, in these cases we also obtain polynomial speed ups over
the best classical algorithms available.

\item Note that if we have an efficient evaluation for $\mathrm{ICCC}%
_{\epsilon }$ then we also have an efficient evaluation for $\mathrm{ICCC}%
_{\epsilon }$, provided $\epsilon >\epsilon ^{\prime }$.

\item We provide a discussion of the computational complexity, both
classical and quantum, in subsection \ref{complexity}. As argued there, we
obtain a polynomial speed up in the difference between the number of edges
and vertices and an exponential speed up in $q$ over the best current
classical algorithm for the $\mathrm{ICCC}_{\epsilon }$ class of graphs.
\end{enumerate}

\begin{mycorollary}
\label{cor} For a given graph $\Gamma $, whose CMM is the direct sum of the
CMMs of two graphs $\Gamma _{1}$ and $\Gamma _{2}$ in $\mathrm{ICCC}%
_{\epsilon }$, a quantum computer will be able to return $Z_{\Gamma }$ with
a running time equal to the sum of the running times required to obtain $%
Z_{\Gamma _{1}}$ and $Z_{\Gamma _{2}}$.
\end{mycorollary}

Proofs of the main theorem and the corollary are provided in Sections \ref%
{proof:th} and \ref{proof:cor}.

\subsection{Background}

\label{back}

Theorem \ref{th 1} connects the problem of estimating the Potts partition
function to a quantum algorithm for Gauss sums, via weight enumerators for
irreducible cyclic codes. In somewhat more detail, the connections we need
are as follows. In \cite{barg}, it was shown that the Potts partition
function can be written as the weight enumerator of the cocycle code of the
graph $\Gamma $, over which the Potts model is defined. Weight enumerators
of irreducible cyclic codes are related to Gauss sums via the McEliece
Theorem \cite{Baumert:72}.

\subsubsection{Cycle Matroid Matrix Representation of a Graph}

A connected component of a graph is any subset of vertices which are all
connected to each other via a path along the graph's edges. We denote the
number of connected components by $c(\Gamma )$. The incidence matrix of a
finite graph $\Gamma (E,V)$ is a $|V|\times |E|$ binary matrix where column $%
c$ represents edge $c$ with non-zero entries in row $i$ and $j$ if and only
if vertices $i$ and $j$ are the boundaries of edge $c$. Every finite graph $%
\Gamma $ also gives rise to a cycle matroid matrix (CMM) \cite{Welsh:matroid}%
, which essentially captures the presence and locations of cycles in the
graph.

\begin{definition}
\label{def:CMM}The cycle matroid matrix of a graph $\Gamma =(E,V)$, CMM($%
\Gamma $), is formed as follows:\ write down the incidence matrix of $\Gamma 
$ using $1$ for the $i$th and $-1$ for the $j$th rows, where $i<j$. Then
apply elementary row operations and Gaussian reduction to obtain a $%
(|V|-c(\Gamma ))\times |E|$ matrix of the form $[I_{|V|-c(\Gamma )}|X]$,
where $I_{a}$ is the $a\times a$ identity matrix and $X$ is a $(|V|-c(\Gamma
))\times (|E|-|V|+c(\Gamma ))$ matrix. This is CMM($\Gamma $) (See Prop.
4.7.14 of \cite{Gross:book}).
\end{definition}

We give more details on cycle matroids in Appendix~\ref{appA}. As an example
consider the square [$|V|=|E|=4$, $c(\Gamma )=1$] and its incidence matrix

\begin{equation*}
\left( 
\begin{array}{cccc}
1 & 0 & 0 & -1 \\ 
-1 & 1 & 0 & 0 \\ 
0 & -1 & 1 & 0 \\ 
0 & 0 & -1 & 1%
\end{array}%
\right) .
\end{equation*}%
Applying elementary row operations and Gaussian reduction one obtains the CMM

\begin{equation*}
\left( 
\begin{array}{cccc}
1 & 0 & 0 & -1 \\ 
0 & 1 & 0 & -1 \\ 
0 & 0 & 1 & -1%
\end{array}%
\right) ,
\end{equation*}%
which is indeed of the form $[I_{|V|-c(\Gamma )}|X]$ with dimensions as in
the definition, i.e., $X$ is $(4-1)\times (4-4+1)$. Over $\mathbf{Z}_{2}$
one would replace all the $-1$'s with $+1$'s. The column space of this
matrix represents the cycle structure of the graph where a cycle (or
circuit) is a path in the graph for which the first vertex of the path is
the same as the last. Any set of columns that are linearly dependent
indicate a cycle. The first three columns in the CMM\ of the square are
linearly independent, but together with the fourth column they become
linearly dependent, since there is a cycle in the graph involving the
corresponding four edges.

What is the equivalence class of graphs with the same CMM? This is answered
by the following:

\begin{definition}
Two graphs $G$ and $G^{\prime }$ are called 2-isomorphic if there exists a $%
1-1$ correspondence between the edges of $G$ and $G^{\prime }$ such that the
cycle (or circuit) relationships are preserved.
\end{definition}

Thus all 2-isomorphic graphs have the same CMM (up to elementary row and
column operations).

\subsubsection{Irreducible Cyclic Codes}

We provide some background material in coding theory which allows us to
exhibit the connection between Eq.~(\ref{newpotts}) and the so-called
cocycle code of the graph $\Gamma $. (A good reference is \cite{Lint:book}.)

\begin{definition}
Let $\mathbf{F}_{q}$ be a finite field with $q$ prime. A linear code $C$ is
a $k$ dimensional subspace of the vector space $\mathbf{F}_{q}^{n}$ and is
referred to as an $[n,k]$ code. The code is said to be of length $n$ and of
dimension $k$.
\end{definition}

In our case $q$ is the the number of possible states per spin.

\begin{definition}
A $k\times n$ matrix whose rows are a basis for $C$ is called a generator
matrix for $C$.
\end{definition}

Recall from Definition~\ref{def:CMM} that CMM($\Gamma $) is a $(|V|-c(\Gamma
))\times |E|$ matrix. The $|E|$ columns of CMM($\Gamma $) reflect the cycle
structure of the given graph via linear independence in the vector space $%
\mathbf{F}_{q}^{n}$ (see Appendix~\ref{appA}). We now view the $|V|-c(\Gamma
)$ rows of the CMM as generating an $[n=|V|,k=|E|-c(\Gamma )]$
\textquotedblleft cocycle code\textquotedblright\ $C$:

\begin{definition}
The cocycle code $C(\Gamma )$ of a graph $\Gamma $ is the row space of CMM($%
\Gamma $). \cite{barg}
\end{definition}

We focus our attention on a subclass known as \emph{cyclic codes} and a
smaller subclass known as \emph{irreducible cyclic codes}.

\begin{definition}
A linear code $C$ is a cyclic code if for any word $(c_{0},c_{1},\dots
,c_{n-1})\in C$, also $(c_{n-1},c_{0},c_{1},\dots ,c_{n-2})\in C$. If $C$
contains no subspace (other than $0$) which is closed under cyclic shifts
then it is irreducible cyclic.
\end{definition}

Cyclic codes have an interesting underlying algebraic structure which we
review in Appendix~\ref{appB}. In general the generator matrix of an $[n,k]$
cyclic code can be written as 
\begin{equation}
\left( 
\begin{array}{cccccccc}
g_{0} & g_{1} & \cdots  & g_{n-k} & 0 & 0 & \cdots  & 0 \\ 
0 & g_{0} & \cdots  & g_{n-k-1} & g_{n-k} & 0 & \cdots  & 0 \\ 
0 & 0 & \cdots  &  &  &  & \cdots  & 0 \\ 
0 & 0 & \cdots  &  & g_{0} & g_{1} & \cdots  & g_{n-k}%
\end{array}%
\right) .
\end{equation}%
The non-zero matrix elements can be used to construct the \textquotedblleft
generator polynomial\textquotedblright\ $g(x)=g_{0}+g_{1}x+\cdots
g_{n-k}x^{n-k}$. Both can be used to generate a cyclic code; the manner in
which this is done via the generator polynomial is reviewed in Appendix~\ref%
{appB}. For an $[n,k]$ non-degenerate irreducible cyclic code (no words are
repeated) the relation between $n$ and $k$ is $k=\mathrm{ord}_{q}n$, i.e., $k
$ is the smallest integer such that $q^{k}=1\, \mathrm{mod}\, n$. Equivalently, $%
q^{k}-1=nN$, where $N$ counts the number of equivalence classes under cyclic
permutations of words, which is an upper bound on the number of different
weights. Non-degenerate irreducible cyclic codes have generator polynomials
of the form 
\begin{equation*}
g_{i}(x)=\frac{x^{n}-1}{w_{i}(x)}
\end{equation*}%
where $x^{n}-1=w_{1}(x)w_{2}(x)\cdots w_{t}(x)$ is the decomposition of $%
x^{n}-1$ into irreducible factors. In this work we consider only
non-degenerate irreducible cyclic codes. Here $t$ is the number of $q$%
-cyclotomic cosets $\textrm{mod}n$ (see subsection \ref{cyclo}).

Next we explain the connection to weight enumerators.

\begin{definition}
Let $C$ be a linear code of length $n$ and let $A_{i}$ be the number of
vectors in $C$ having $i$ non-zero entries (Hamming weight of $i$) . Then
the weight enumerator of $C$ is the bi-variate polynomial%
\begin{equation*}
A(x,y)=\sum_{i=0}^{n}A_{i}x^{n-i}y^{i}.
\end{equation*}%
The set $\{A_{i}\}$ is called the weight spectrum of the code.
\end{definition}

In this paper our only concern for the weight spectrum is its connection to
the Potts partition function, but in coding theory it can be used to reveal
information about the effiency of a code \cite{Lint:book}. The connection
between equation (\ref{newpotts}) and the cocycle code of the graph $\Gamma $
for the Potts model is given in the following theorem proved in \cite{barg}.

\begin{theorem}
\label{barg} Let $A(x,y)$ be the weight enumerator of the $%
[n=|E|,k=|V|-c(\Gamma )]$ cocycle code $C(\Gamma )$ of the graph $\Gamma
=(E,V)$, and let the number of states per spin (vertex) in the corresponding
Potts model be a prime $q$. Then 
\begin{equation}
Z_\Gamma (y)=y^{-n}q^{c(\Gamma )}A(1,y).  \label{pottsA}
\end{equation}
\end{theorem}

We take $q$ to be prime and not a power of a prime to simplify matters. In
this manner the cocycle code has words whose entries are in $\mathbf{F}_{q}$%
, as will the corresponding irreducible cyclic code in the trace
representation over $\mathbf{F}_{q^{r}}$ -- see Appendix~\ref{appB} for
details.

The connection between the Potts partition function and weight enumerators
can also be understood via a previous result which shows that $Z$ is
equivalent to the Tutte polynomial (under certain restrictions) and that the
weight polynomial of a linear code is also equivalent to the Tutte
polynomial \cite{tuttepotts}. We also note that a relation similar to Eq.~(%
\ref{pottsA})\ was established in \cite{Lidar:QWGT} for the Ising spin glass
partition function and so-called quadratically signed weight enumerators,
along with a discussion of computational complexity.

\subsection{Testing the graph for membership in the $\mathrm{ICCC}_{\protect%
\epsilon }$\ class}

\label{testing}

We now have the tools to address the issue of whether a graph should be
accepted as input into the main algorithm, i.e., whether a graph belongs to
the $\mathrm{ICCC}_{\epsilon }$ \ class. This is handled as follows.

\noindent - \texttt{Input:} A graph $\Gamma $ with $|E|$ edges and $|V|$
vertices, the given Galois field of $q^{k}$ elements and $\epsilon $. 

\noindent - \texttt{Output:} Accept or Reject. Let $n=|E|$ and $%
k=|E|-|V|+c(\Gamma )$ as in the main theorem.

\noindent - \texttt{Overall Complexity:} $O(|E|\cdot k^{2}\log k\log \log k)$
due the ability to take the discrete log $|E|$ times 
efficiently with a quantum computer \cite{Shor:97}.

\begin{enumerate}
\item Compute $\theta_{n,k}$ as given in Definition \ref{iccc}.

\item Find CMM($\Gamma $). It is a $(n-k)\times n$ matrix of the form $%
[I_{n-k}|X]$, where $X$ is a $(n-k)\times k$ matrix. Form the $k\times n$
(transpose parity check) matrix $H^{T}=[-X^{T}|I_{k}]$. $H^{T}$ generates an 
$[n,k]$ code $C^{\bot }(\Gamma )$ that is dual to the cocycle code $C(\Gamma
)$.

\item Determine if $\epsilon \leq \frac{q^{\theta _{n,k}-1}}{4\sqrt{q^{k}}}$
and if $k$ is the multiplicative order of $q \, \mathrm{mod}\, n$ (i.e., $k$ is the
smallest integer such that $q^{k}=1\, \mathrm{mod}\, n$). If both are true then go
to the next step. Else skip the next step and continue.

\item Main Loop:

\begin{enumerate}
\item Fix a basis of $GF(q^{k})$ over $GF(q)$ and consider the columns of $%
H^{T}$ as coordinate vectors of some elements $g_{i}$ of $GF(q^{k})$.

\item Calculate the discrete logarithms $\log (g_{i})$ of each $g_{i}$ with
respect to a fixed primitive element $g$ (every element in the field can be
written as $g^{l}$ for some $l$) of $GF(q^{k})$ using Shor's algorithm \cite%
{Shor:97} on a quantum computer.\footnote{%
For every non-zero $x\in \mathbf{F}_{q^{r}}/\{0\}$ the discrete logarithm
with respect to a primitive element (i.e., generator) $g$ of $\mathbf{F}%
_{q^{r}}$ is given by $\log _{g}(x)=\log _{g}(g^{j})=j\, \mathrm{mod}\, (q^{r}-1).$%
}

\item Accept or Reject $\Gamma $ based on the fact that $C^{\bot }$ is
(equivalent to) an irreducible cyclic code if and only if the numbers $\log
(g_{i})$ are are some permuted list of consecutive integer multiples of $%
N:=(q^{k}-1)/n$ in some order. This is due to the fact that by definition
the generator matrix of an irreducible cyclic code is equivalent to $(1$ $%
g^{Nj}$ $g^{2Nj}$ $...~g^{(n-1)Nj})$ where $\gcd (n,j)=1$ \cite{Lint:book}.
\end{enumerate}

\item Step c failed. Using elementary row operations transform $H^{T}$ to a
block diagonal matrix if possible. If not possible then Reject. If possible
then go to step c and input each sub-matrix and continue.
\end{enumerate}

\subsection{Proof of the Main Theorem}

\label{proof:th}


\subsubsection{Preliminaries}

We introduce Gauss sums as this is the vital link between the Potts
partition function and quantum computation. Appendix~\ref{appC} contains a
more detailed exposition as well as an outline of a proof that there exists
a polylog quantum algorithm for estimating Gauss sums. Here however it is
essential only to understand the following. Given a field $\mathbf{F}%
_{q^{k}} $, there is a multiplicative and additive group associated with it.
Namely, the multiplicative group is $\mathbf{F}_{q^{k}}^{\ast }=\mathbf{F}%
_{q^{k}}\setminus 0$ and the additive group is $\mathbf{F}_{q^{k}}$ itself.
Associated with each group are canonical homomorphisms from the group to the
complex numbers, named the additive and multiplicative characters. The
multiplicative character $\chi $ is a function of the elements of $\mathbf{F}%
_{q^{k}}^{\ast }$ and the additive character is a function of $\mathbf{F}%
_{q^{k}}$ and is parameterized by $\beta \in \mathbf{F}_{q^{k}}$. A Gauss
sum is then a function of the field $\mathbf{F}_{q^{k}}$, the multiplicative
character $\chi $ and the parameter $\beta $, and can always be written as 
\begin{equation}
G_{\mathbf{F}_{q^{k}}}(\chi ,\beta )=\sqrt{q^{k}}e^{i\gamma },  \label{gauss}
\end{equation}%
where $\gamma $ is a function of $\chi $ and $\beta $. It is in general
quite difficult to find the angle $\gamma $. The complexity of estimating
this quantity via classical computation is not known but there is evidence
that it is hard \cite{van}.

We now introduce the trace function over finite fields.

\begin{definition}
Let $q$ be prime, $k$ a positive integer, and let $\mathbf{F}_{q^{k}}$ be
the finite field with $q^{k}-1$ non-zero elements. The trace is a mapping $%
\mathrm{Tr}:\mathbf{F}_{q^{k}}\mapsto \mathbf{F}_{q}$ and is defined as
follows. Let $\xi \in \mathbf{F}_{q^{k}}$. Then%
\begin{equation}
\mathrm{Tr}(\xi )=\sum_{j=0}^{k-1}\xi ^{q^{j}}.  \label{eq:Tr}
\end{equation}
\end{definition}

Now let $q^{k}=1+nN$ where $n$ and $N$ are both positive integers, and let $%
\gamma $ generate the multiplicative (cyclic) group $\mathbf{F}%
_{q^{k}}^{\ast }=\mathbf{F}_{q^{k}}\backslash \{0\}$. Each of the $q^{k}$
words of an $[n,k]$ irreducible cyclic code may then be uniquely associated
with an element $x\in \mathbf{F}_{q^{k}}$ and may be written as

\begin{equation}
(\mathrm{Tr}(x),\mathrm{Tr}(x\gamma ^{N}),\mathrm{Tr}(x\gamma ^{2N}),\dots ,%
\mathrm{Tr}(x\gamma ^{(n-1)N})),  \label{word}
\end{equation}%
where $k$ is the smallest integer such that $q^{k}=1\, \mathrm{mod}\, n$. For a
proof of this statement see \cite{Berndt:book} or \cite{Moiso:97}.

As stated in Theorem \ref{th 1}, we are essentially interested in obtaining
the weight spectrum of $[n,k]$ irreducible cyclic codes. The number of words
with different non-zero weight is at most $N$ where $N=(q^{k}-1)/n$ (for a proof see
Proposition \ref{prop:N} in Appendix \ref{appB}). Now let $w(x)$ be the
Hamming weight of the code word associated with $x\in \mathbf{F}%
_{q^{k}}^{\ast }$. The McEliece Theorem connects the weights of words of
irreducible cyclic codes to Gauss sums.

\begin{theorem}
\label{mceliece} \textbf{(McEliece Formula)} Let $w(y)$ for $y\in \mathbf{F}%
_{q^{k}}^{\ast }$ be the weight of the code word given by Eq.~(\ref{word}),
let $q^{k}=1+nN$ where $q$ is prime and $k$, $n$ and $N$ are positive
integers, let $d=\mathrm{gcd}(N,(q^{k}-1)/(q-1))$, and let the
multiplicative character $\bar{\chi}$ be given by $\bar{\chi}(\gamma )=\exp
(2\pi i/d)$, where $\gamma $ generates $\mathbf{F}_{q^{k}}^{\ast }$. ($\bar{%
\chi}$ is called the character of order $d$.) Then the weight of each word
in an irreducible cyclic code is given by 
\begin{equation}
w(y)=\frac{q^{k}(q-1)}{qN}-\frac{q-1}{qN}\sum_{a=1}^{d-1}\bar{\chi}%
(y)^{-a}G_{\mathbf{F}_{q^{k}}}(\bar{\chi}^{a},1).  \label{eq:Mc}
\end{equation}
\end{theorem}

For a proof of this see \cite{Berndt:book}.

The important feature here is that if we had the ability to efficiently
estimate $G_{\mathbf{F}_{q^{k}}}(\chi ,\beta )$, then we would be able to
find the weights of the words in an irreducible cyclic code efficiently
under the restrictions mentioned in Theorem \ref{th 1} . This would in turn
allow us to find the weight spectrum $\{A_{i}\}$ of the code. The following
theorem reveals that a quantum computer can efficiently approximate Gauss
sums.

\begin{theorem}
\textbf{{(van Dam \& Seroussi \cite{van})}} For any $\epsilon >0$, there is
a quantum algorithm that estimates the phase $\gamma $ in $G_{\mathbf{F}%
_{q^{k}}}(\chi ,\beta )=\sqrt{q^{k}}e^{i\gamma }$, with expected error $%
E(\lvert \gamma -\tilde{\gamma}\rvert )<\epsilon $. The time complexity of
this algorithm is bounded by $O(\frac{1}{\epsilon }\cdot (\log (q^{k}))^{2})$%
. \cite{van}
\end{theorem}

(For details see Appendix \ref{appC}.)

The Gauss sum algorithm allows one to estimate $\gamma $ in Eq. (\ref{gauss}%
) to within any accuracy $\epsilon $, i.e., the algorithm returns $\gamma
^{\prime }$ such that $|\gamma ^{\prime }-\gamma |<\epsilon $. The hope is
that if one can approximate $\gamma $ precisely enough then one would get an
exact evaluation of the weight. In fact an essential step here is to use a
quantum computer to obtain a list of approximate angles $\{\gamma
_{t}^{\prime }\}$ for $t=1,\dots ,d-1$ for $d$ given above.

The next theorem gives some minimum distance between weights so that we can
choose an appropriate error that will allow one to be able to distinguish
between weights, which allows us to obtain accurate coefficients for $A(1,y)$ and
hence exact values for the exponents.

\begin{theorem}
\label{mceliece2} \textbf{{(McEliece \cite{aubry})}} All the weights of an $%
[n,k]$ irreducible cyclic code are divisible by $q^{\theta_{n,k} -1}$, where 
$\theta_{n,k}$ is given in Definition \ref{iccc}.
\end{theorem}

\subsubsection{The Proof}

We are now ready to prove Theorem \ref{th 1}.

\begin{proof}
Assume that a given graph $\Gamma =(E,V)$ is a member of $\mathrm{ICCC}%
_{\epsilon }$, where $n=|E|$ and $k=|E|-|V|+c(\Gamma )$. Hence it is given
that $\epsilon \leq \frac{q^{\theta _{n,k}-1}}{4\sqrt{q^{k}}}\equiv \epsilon
_{0}$. We want to obtain $Z_{\Gamma }$ for either the fully ferromagnetic or
anti-ferromagnetic Potts model. It follows from Definition \ref{iccc} that
the dual of the cocycle code of $\Gamma $ is an irreducible $[n,k]$ cyclic
code. We must demonstrate that we can obtain the weight enumerator $A(1,y)$
of this dual code within the claimed number of steps. As mentioned above,
since $nN=q^{k}-1$ there are then at most $N$ different weights, with at
least $n$ words of each weight (see the appendix). In order to find the
spectrum $\{A_{i}\}$, we are faced with the computational task of finding
the range of 
\begin{equation}
S(i)=\frac{q^{k}(q-1)}{qN}-\frac{q-1}{qN}\sum_{a=1}^{d-1}\bar{\chi}(\alpha
^{i})^{-a}\sqrt{q^{k}}e^{i\widetilde{\gamma _{a}}}  \label{S(i)}
\end{equation}%
(where again $d=\gcd (N,q^k-1/q-1)$ and $i\in \{0,\dots ,N-1\}$) and
then performing a tally.

The proof consists of five main parts:

\begin{enumerate}
\item Proof that with $\epsilon $ bounded by $\epsilon _{0}$ as given, it is
possible to distinguish between weights of the words of the code that
corresponds to the given graph. This ability allows for an exact evaluation
of $Z_{\Gamma }$.

\item We need to justify our asymptotic approach and show that for a fixed
error $\epsilon $ there are a countable number of graphs in $\mathrm{ICCC}%
_{\epsilon }$.

\item Proof that the success probability $\delta $ is as stated in the
Theorem.

\item Proof that the running time is as stated in the Theorem.

\item A transformation from the dual (irreducible cyclic)\ code to the
cocycle code of the graph whose Potts partition function we are evaluating.
\end{enumerate}

Let us now prove each of these five parts.

1. The first question we must address is the following:\ how small do we
need to make the error $\epsilon $ in the phases returned in the Gauss sum
approximation algorithm so that we will be able to distinguish between
weights? We now show that $\epsilon \leq \epsilon _{0}$ is sufficient, and
hence that for every member of the class ICCC$_{\epsilon }$ it is possible
to distinguish between weights.

Let $\widetilde{w(y)}$ be the approximated weight returned by the quantum
Gauss sum algorithm. It follows from Theorem \ref{mceliece2} that two
consecutive weights are separated by a distance that is an integer multiple
of $q^{\theta _{n,k}-1}$. Hence, a sufficient condition for being able to
associate $\widetilde{w(y)}$ with the correct weight $w(y)$ (and not another
neighboring weight) is: 
\begin{equation}
|w(y)-\widetilde{w(y)}|<\frac{q^{\theta _{n,k}-1}}{2}.  \label{wtdiff}
\end{equation}%
Let the error between the actual phase $\gamma _{i}$ and the approximated
phase $\widetilde{\gamma _{i}}$ be $\epsilon $, i.e.,%
\begin{equation*}
|\gamma _{i}-\widetilde{\gamma _{i}}|<\epsilon .
\end{equation*}%
Let us derive a bound on $\epsilon $. Taking $w(y)$ given in Theorem \ref%
{mceliece} and the necessary bound given in equation (\ref{wtdiff}) we find
that we need the inequality 
\begin{equation}
\left\vert \sum_{a}\bar{\chi}(y)^{-a}e^{i\gamma _{a}}-\sum_{a}\bar{\chi}%
(y)^{-a}e^{i\widetilde{\gamma _{a}}}\right\vert <\frac{qN}{(q-1)}\frac{%
q^{\theta _{n,k}-1}}{2}\frac{1}{\sqrt{q^{k}}}  \label{wtdiff2}
\end{equation}%
to be satisfied. Now, we have 
\begin{eqnarray*}
\left\vert \sum_{a}\bar{\chi}(y)^{-a}e^{i\gamma _{a}}-\sum_{a}\bar{\chi}%
(y)^{-a}e^{i\widetilde{\gamma _{a}}}\right\vert &\leq &\sum_{a}\left\vert
e^{i\gamma _{a}}-e^{i\widetilde{\gamma _{a}}}\right\vert \\
&\leq &\sum_{a}\left( \left\vert \cos (\gamma _{a})-\cos (\widetilde{\gamma
_{a}}))\right\vert +\left\vert (\sin (\gamma _{a})-\sin (\widetilde{\gamma
_{a}}))\right\vert \right) \\
&\leq &2(d-1)\left\vert \gamma _{a}-\widetilde{\gamma _{a}}\right\vert
<2(d-1)\epsilon
\end{eqnarray*}%
where the last inequality follows from the Mean Value Theorem of elementary
calculus. Therefore, if we impose%
\begin{equation}
\epsilon <\frac{qN}{(q-1)(d-1)}\frac{q^{\theta _{n,k}-1}}{4}\frac{1}{\sqrt{%
q^{k}}}  \label{eps1}
\end{equation}%
then inequality (\ref{wtdiff2}) is satisfied. Consider the factor $\frac{qN}{%
(q-1)(d-1)}$. Noting from $N\leq \alpha k^{s}$ that $d=\mathrm{gcd}%
(N,(q^{k}-1)/(q-1))\leq \alpha k^{s}=N$, it follows that $1<\frac{qN}{%
(q-1)(d-1)}\leq \frac{q}{q-1}N=O(k^{s(k)})$. Thus we can replace the bound (%
\ref{eps1}) by the tighter bound 
\begin{equation}
\epsilon <\frac{q^{\theta _{n,k}-1-k/2}}{4}=\epsilon _{0},  \label{eps2}
\end{equation}%
and this $\epsilon $ is definitely small enough to satisfy the required
bound given in Eq. (\ref{wtdiff}).\footnote{%
Note, however, that when $d=2$, we have in fact $\epsilon <k^{s(k)}\epsilon
_{0}$ where $\epsilon _{0}=\frac{q^{\theta _{n,k}-1}}{4\sqrt{q^{k}}}$, and
in this case the computational cost of the algorithm (see Theorem \ref{th 1}%
) is scaled down from $O(\frac{1}{\epsilon _{0}}k^{2s(k)}(\log q)^{2})$ to $%
O(\frac{1}{\epsilon _{0}}k^{s(k)}(\log q)^{2})$, where the upper bound (\ref%
{eps2})\ still applies. This means that within the family ICCC$_{\epsilon
_{0}}$ some instances can be solved faster than others by a factor of $%
k^{s(k)}$, at fixed $\epsilon _{0}$.} Hence, if $\epsilon <\epsilon _{0}$ it
is possible to resolve the weights $\widetilde{w(y)}$ for different words $y$%
. This, in turn, gives us the ability to exactly reconstruct the weight
enumerator $A$, and from there the partition function $Z_{\Gamma }$.

2. We prove the following lemma.

\begin{lemma}
Given a fixed $\epsilon <1$ there are countably many graphs in $\mathrm{ICCC}%
_{\epsilon }$, i.e., there are infinitely many corresponding irreducible
cyclic codes $[n_{i},k_{i}]$ such that $\{\theta _{n_{i},k_{i}}\}$ satisfies 
\begin{equation}
\frac{1}{4}q^{\theta _{n_{i},k_{i}}-1-\frac{k_{i}}{2}}>\epsilon .
\label{zero}
\end{equation}
\end{lemma}

What this means is that there is at least one family of graphs for a given
fixed $\epsilon $ for which one will be able to obtain the exact Potts
partition function. This also justifies the complexity arguments used herein.

\begin{proof}
We shall construct one such family and show that it satisfies the required
relations. For simplicity take $\epsilon =4\epsilon $. We must construct one
family of graphs for which the corresponding irreducible cyclic codes, $%
[n_{i},k_{i}]$, satisfy 
\begin{equation}
\theta _{n_{i},k_{i}}+\log _{q}(\epsilon ^{-1})>1+\frac{k_{i}}{2}.
\label{one}
\end{equation}%
Take $q$ fixed and consider the following countable set of irreducible
cyclic codes: $\{[q^{m}-1,k_{m}]\}_{m=1,2,3,\dots }$. First we must note that%
\begin{equation}
\theta _{q^{m}-1,k_{m}}=m.  \label{two}
\end{equation}%
This follows from the properties of addition in base $q$:$\ $the $k$ digits
of $q^{k}-1$ in base $q$ are all $(q-1)$, and adding integer multiples of $%
q^{k}-1$ will not decrease the digit sum. I.e.,%
\begin{equation*}
S^{\prime }(\eta (q^{m}-1))\geq m(q-1)\quad \forall \eta \in \mathbb{N}.
\end{equation*}%
This is important to keep in mind when we consider extending this family
later in this proof. Now we must demonstrate that there is at least one $%
k_{m}$ that satisfies Eq. (\ref{one}). Because we are dealing with
irreducible cyclic codes we must have 
\begin{equation*}
q^{k_{m}}=1\textrm{mod}n=1\textrm{mod}(q^{m}-1).
\end{equation*}%
This is trivially satisfied by $k_{m}=m$ and indeed Eq. (\ref{one}) becomes $%
m-\frac{m}{2}>1+\log _{q}{\epsilon }$, which is clearly true for any $%
\epsilon <1$. This family is computationally trivial, however, being
that $N=1$.

Let us now extend this family to include many interesting instances. Let us
first consider a fixed code $[q^{m}-1,k_{m}]$ (i.e., $N=1$, $m$ fixed). Let
us next generate a family of codes $\{[\eta
_{j}(q^{m}-1),k_{mj}]\}_{j=1,2,\dots ,M}$ by taking integer multiples $\eta
_{j}$ of $q^{m}-1$, and picking $k_{mj}\geq k_{m}$ such that $\eta
_{j}(q^{m}-1)N=q^{k_{mj}}-1$ (this is just the irreducible cyclic code
condition $nN=q^{k}-1$). We obtain a finite set of codes ($M<\infty $)
because it follows from Eq. (\ref{zero}) that eventually the $\{k_{mj}\}_{j}$
will become too large for the fixed error $\epsilon $, for each $m$. We then
do this for every $m\in \mathbf{N}$ paying special attention to the integer
multiples $\eta _{j}$. The $\eta _{j}$ are selected in this construction so
that two conditions are satisfied:\ (i) the corresponding $\{k_{mj}\}_{j}$
are sufficiently small to ensure that Eq. (\ref{zero}) is satisfied, (ii)
that $N$ is bounded by $\{O(k_{mj}^{s})\}_{m,j}$.

Regarding (i), the steps above are conveniently summarized as the following
loop:

Given $\epsilon $:

\begin{enumerate}
\item For $m=1,2,...$

\begin{enumerate}
\item[2.] Repeat $j=1,2,...$

$n:=j(q^{m}-1)$

calculate $k_{mj}=$ord$_{q}(n)$

if $q^{\theta _{n,k_{mj}}-1-k_{mj}/2}<\epsilon $ then reject $k_{mj}$, else
accept $k_{mj}$ and let $\eta _{j}\equiv j$.

Until $j=M$
\end{enumerate}
Until enough instances were found.

\end{enumerate}

Note that we are guaranteed to find such a non-empty finite set $%
\{k_{mj}\}_{j}$ due to the fact that if $\gcd (q,\eta _{j}(q^{m}-1))=1$,
then $\exists k_{mj}\in \mathbf{N}$ such that $q^{k_{mj}}=1\, \mathrm{mod}\,  j(q^{m}-1)$
(see, e.g., Th. 7-1 of \cite{Andrews:book}).

Regarding (ii), we still need to show that there exist solutions $N_{mj}$ to 
$q^{k_{mj}}-1=nN_{mj}$ that scale as $O(k_{mj}^{s})$. To see why such
solutions exist consider solving $q^{k}-1=nN$ with $N=\alpha k^{s}$ and $%
n=\eta (q^{m}-1)$, where $\alpha \in \mathbf{R}$ (we have dropped the
subscripts for simplicity). The solution is 
\begin{equation}
m=\log _{q}[(q^{k}-1)/(\alpha \eta k^{s})+1].  \label{eq:m}
\end{equation}%
In the loop above, only those $m$'s satisfying Eq. (\ref{eq:m}) are
acceptable in terms of the scaling of our algorithm. However, note that
asymptotically Eq. (\ref{eq:m}) yields $m=k-s\log _{q}k-\log _{q}\alpha \eta 
$. This means that for every value of $k$ and $s$ it is possible to adjust $%
\alpha $ such that $m$ is an integer by letting $s\log _{q}k=\log _{q}\alpha
\eta $.

At this point we have constructed an infinite family of pairs $[n,k_{mj}]$
[where $n=j(q^{m}-1)$ and where $m$ satisfies Eq. (\ref{eq:m})], each of
which defines a graph which is a member of the set ICCC$_{\epsilon }$.
\end{proof}

Finally, we should mention without proof, that one can \textquotedblleft
fill\textquotedblright\ this family of graphs by considering the multitude
of cases which do not conform to the restrictions in this construction, but
which do obey relation (\ref{one}) and obey the asymptotic conditions given
in definition (\ref{iccc}). Moreover, the graphs we have constructed are
quite sparse but they are only a subset of $\mathrm{ICCC}_{\epsilon }$.
There are many more interesting graphs that can be handled by this fixed
error bound. For example, graphs which are the direct sum of many copies of
a smaller graph are excluded from this family. Further one may accept an
error $\epsilon $ that decreases polynomially in $k$ for example and define
a family of graphs in that way. We do not pursue this here.

3. In the van Dam-Seroussi algorithm (Theorem 1 in \cite{van}), a prepared
state must must go through a phase estimation. In \cite{Nielsen:book} it is
demonstrated that if the number of qubits used in phase estimation is $%
t=\log 1/\epsilon +\log (2+1/(2\delta ))$ then the probability of success is
at least $1-\delta $. Ref. \cite{van2}[p.7] states that for the Gauss sum
algorithm $t=2\log (q^{k}-1)$. After some elementary algebra we obtain $%
\delta =[2((q^{k}-1)^{2}\epsilon -2]^{-1}$. By the Chernoff bound, for fixed
problem size $k$, we only need to pick $\epsilon $ such that the probability
of failure $\delta $ is less than $1/2$.

4. (a) We have that if $\alpha $ is a generator for $\mathbf{F}%
_{q^{k}}^{\ast }$ and if $i=j\, \mathrm{mod}\, n$, then the code words associated
with $\alpha ^{i}$ and $\alpha ^{j}$ are cyclic permutations of each other,
and therefore are of the same weight. Let us denote by $[\alpha ^{i}]$ the
(equivalence) class of all words $\{\alpha ^{j}\}_{j}$ with $i=j\, \mathrm{mod}\, %
n $. In this step we wish to find the weight of $[\alpha ^{i}]$. This weight
is given by 
\begin{equation}
S(i)=\frac{q^{k}(q-1)}{qN}-\frac{q-1}{qN}\sum_{a=1}^{d-1}\bar{\chi}(\alpha
^{i})^{-a}\sqrt{q^{k}}e^{i\widetilde{\gamma _{a}}}.
\end{equation}%
Hence (up to irrelevant classical computations) the computational cost of
computing $S(i)$ is $d-1$ times the cost of computing $\widetilde{\gamma _{a}%
}$. For any graph in $\mathrm{ICCC}_{\epsilon }$, obtaining these $d-1$
phases has a (quantum) cost of $O(dk^{2}(\log q)^{2})$, where $d$ is bounded
above by $N$. This comes from the complexity of computing the Gauss sum $d$
times. (Recall that one has to repeat this algorithm $1/\epsilon $ times in
order to ensure that we obtain a sufficiently close approximation.)

(b) How many times must we compute $S(i)$? The number of times is the number
of different equivalence classes $\{[\alpha ^{i}]\}$. Each equivalence class 
$[\alpha ^{i}]$ is clearly of size $n$, and there are $q^{k}-1$ words.
Recall that $nN=q^{k}-1$ for non-degenerate irreducible cyclic codes, and
hence $N$ is the number of different equivalence classes. (Actually the
answer to \textquotedblleft How many times must we compute $S(i)$%
?\textquotedblright\ is that one must only do this for the number of
cyclotomic cosets of $N$ -- see subsection \ref{cyclo}).

(c) For given $S(i)$ we must compute a sum over $d$ terms. The cost of
computing each such term is constant once we have obtained the phases $%
\widetilde{\gamma _{a}}$ [which we have, in step (a)]. Combining this with
step (b), we see that the total cost of computing all $S(i)$'s is $(d-1)N$.

At this point the total computational cost is therefore $\max [O(dk^{2}(\log
q)^{2}),O(dN)]$. We choose $N=O(k^{s(k)})$ so if one takes $s(k)$ to be a
constant, then the algorithm is polynomial in $k$. Thus, the overall time
complexity is $O(d\cdot k^{\max [2,s(k)]}(\log q)^{2})$. Being that $d\leq
N=O(k^{s(k)})$, the complexity is ultimately $O(k^{2\max [1,s(k)]}(\log
q)^{2})$.

(d) We now have the list $\{S(i)\}$. Next, a tally of all the weights has to
be done which has complexity $O(k^{s(k)/2})$ using quantum counting \cite%
{count}. The tally will return all the weights and counts of each weight
(see Section \ref{summary}) which are the exponents and coefficients
respectively, of the polynomial $A(1,y)$ which is the weight enumerator of
the dual of the cocycle code. Note that this step does not effect the
overall complexity of the algorithm as it has a smaller running time then
the previous steps.

5. Note that so far we have dealt with the $[n,k]$ irreducible cyclic code
that is the dual of the cocycle code of $\Gamma $, i.e., we have used $n=|E|$
and $k=|E|-|V|+c(\Gamma )$. However, recall that $\Gamma =\Gamma (E,V)$ and
hence corresponds to the $[n,n-k]=[|E|,|V|-c(\Gamma )]$ code, i.e., the
cocycle code of the graph $\Gamma $ as desired. (This correspondence means
that we can obtain information about interesting graphs by considering codes
of smaller dimension.) Thus, in order to complete the proof we need the
weight enumerator of the $[n,n-k]$ cocycle code itself, so that we can apply
Theorem \ref{barg}. The relation between the weight enumerator $A$ of a code 
$C$ over the field $\mathbf{F}_{q^{k}}$, and the weight enumerator $A^{\bot
} $ of the dual code $C^{\bot }$ is given by the MacWilliams Theorem \cite%
{Lint:book}: 
\begin{equation}
A^{\bot }\left( 1,x\right) =q^{k(k-n)}\left( 1+(q^{k}-1)x\right) ^{n}A(1,y),
\label{mac}
\end{equation}%
where%
\begin{equation}
x\equiv \left( \frac{1-y}{1+(q^{k}-1)y}\right) .
\end{equation}%
Applying the MacWilliams theorem and Barg's theorem [specifically Eq.~(\ref{pottsA}) to $A^{\bot }\left( 1,x\right) $], we arrive at the partition
function%
\begin{equation}
Z\left( x\right) =x^{-n}q^{c(\Gamma )}A^{\bot }\left( 1,x\right) .
\end{equation}%
Recall that $y=e^{-\beta J}$ (where $\beta =\frac{1}{k_{B}T}$); thus we have
the following final expression for the partition function as a function of $%
\beta $:%
\begin{equation}
Z\left( x(\beta )\right) =q^{c(\Gamma )+k(k-n)}\left[ (q^{k}-1)+x(\beta
)^{-1}\right] ^{n}A(1,y(\beta )).  \label{Zfinal}
\end{equation}%
It is simple to verify that given any temperature $T\geq 0$, and for both
positive and negative $J$, $Z\left( x(\beta )\right) $ is always positive,
as it should be.
\end{proof}

\subsection{Proof of the Corrolary}

\label{proof:cor}

We now give the proof of Corrolary \ref{cor}.

\begin{proof}
Assume that we are given a graph $\Gamma (E,V)$ whose CMM is the direct sum
of the CMMs of two graphs $\Gamma _{1}$ and $\Gamma _{2}$ in $\mathrm{ICCC}%
_{\epsilon }$ (we call such a graph $\Gamma $ a \textquotedblleft composite
graph\textquotedblright ). Let $C$ be the code that corresponds to the graph 
$\Gamma $, i.e., $C$ is the cocycle code of $\Gamma $. Let $C_{1}$ and $%
C_{2} $ be the corresponding cocycle codes of $\Gamma _{1}$ and $\Gamma _{2}$%
. This means that we may apply our algorithm to each of these sub-graphs and
obtain their weight enumerators. To do this we need to obtain the weight
enumerators of $C_{1}$ and $C_{2}$ which we can do efficiently. By
definition $C=C_{1}\oplus C_{2}$. If the respective lengths and dimensions
of $C_{1}$ and $C_{2}$ are $[m,l]$ and $[m^{\prime },l^{\prime }]$, then $C$
is an $[m+m^{\prime },l+l^{\prime }]$ linear code and its weight enumerator
will be $W=W_{1}W_{2}$ \cite{Lint:book}. Thus, once one obtains the weight
enumerators of the sub-graphs, one has the weight enumerator of $\Gamma $
and by using the arguments already outlined one can see that we can
efficiently compute $Z_{\Gamma }$.
\end{proof}

The above corollary allows the scheme outlined in this paper to be
efficiently applied to many graphs because if one knows the generator
matrices for $C_{1}$ and $C_{2}$ then one can efficiently construct the
generator matrix for $C$ by just taking the direct sum of the matrices. This
gives a way of constructing examples of graphs for which the Potts partition
function can be efficiently approximated. On the other hand (recall
subsection \ref{testing}), we can efficiently check if a generator matrix
decomposes into a direct sum of smaller matrices and we can efficiently
check if these matrices generate codes whose duals are irreducible cyclic.

\subsection{Reducing the Computational Cost of the Algorithm via Permutation
Symmetry}

\label{cyclo}

We now briefly review the concept of $p$-cyclotomic cosets. As an
introduction see \cite{Lint:book}. Consider the set of integers $%
\{0,1,2,\dots ,N-1\}$ and take $p$ a prime number such that $p$ does not
divide $N$. The $p$-cyclotomic cosets of this set are given by the
collection of subsets 
\begin{equation*}
\{0\},\{1,p,p^{2},\dots ,p^{r(1)}\},\dots ,\{j,jp,jp^{2},\dots
,jp^{r(j)}\},...
\end{equation*}%
where $j$ is an integer and $r(j)$ is the smallest integer such that $%
j(p^{r(j)}-1)=0\, \mathrm{mod}\, N$, i.e., $r(j)$ is the smallest integer before
one begins to get repeats in the coset indexed by $j$. The number of cosets
is finite so $j$ is finite. As an example consider $N=16$ and $p=3$. One
obtains 
\begin{equation*}
\{0\},\{1,3,9,11\},\{2,6\},\{4,12\},\{5,15,13,7\},\{8\},\{10,14\}.
\end{equation*}

With regards to the scheme presented in this paper, we take $q$-cyclotomic
cosets. We are guaranteed that $\gcd (N,q)=1$, which ensures that in our
case the cyclotomic cosets are disjoint. That $\gcd (N,q)=1$ is due to the
fact that there are solutions $x,y\in \mathbf{Z}$ to $Nx+qy=1$ (Thm. 2-4 of 
\cite{Andrews:book}). For example, since $N=\frac{q^{k}-1}{n}$, one can take 
$x=n(q-1)$ and $y=1+q^{k-1}-q^{k}$, which are both integers. The relevance
of the $q$-cyclotomic cosets of $\{0,\dots ,N-1\}$ is that each element in a
given coset has the same value of $S(i)$. This is because of the fact that
the mapping $x\mapsto x^{q}$ is a permutation of $\mathbf{F}_{q^{k}}$ and
that the additive characters obey the identity $\exp (2\pi i$Tr$%
(b^{q})/q)=\exp (2\pi i$Tr$(b)/q)$ for all $b\in \mathbf{F}_{q^{k}}$. Hence $%
S(i)$ is invariant under the mapping $x\mapsto x^{q}$. (See Appendix \ref%
{app:Characters} for details on additive characters and the trace function
Tr.) Therefore we only have to evaluate $S(i)$ for one $i$ in each coset.
The computational cost of computing the coset representatives and the number of
elements in each coset is linear in $N$ \cite{Geraci-vanBussel:07}. This has
the potential of significantly speeding up the algorithm but how much will
clearly depend on the number of cosets generated by each instance. The
number of cosets is given by \cite{Lidl:97} 
\begin{equation}
N_{C}=\sum_{f|N}\frac{\phi (f)}{\mathrm{ord}_{q}f}  \label{N_C}
\end{equation}%
where $\phi (f)$ is the Euler totient (the number of positive integers which
are relatively prime to $f$ and $s=\mathrm{ord}_{q}f$ means that $s$ is the
smallest positive integer such that $q^{s}=1\textrm{mod}f$). Note that $N_{C}$
replaces $N$ in the overall computational cost of our algorithm and $%
N_{C}\leq N$. While this can lead to a significant speedup in some cases,
for the sake of simplicity and of having uniform bounds we will not pursue
this further here.

As an illustration of the power of using cyclotomic cosets, consider the
following numerical example. Let $q=2$, $1/\epsilon \geq 8192$, and consider
a binary $[113,85]$ code which is the dual to a binary $[113,28]$
irreducible cyclic code (i.e., 28 is the smallest integer such that $2^{28}=1%
\textrm{mod}113$). This corresponds to either the fully ferromagnetic or fully
anti-ferromagnetic Ising model on a graph with 113 edges and 86 vertices.
Now note, that $nN=2^{28}-1$ which implies that $N=2375535$. Without the use
of cyclotomic cosets this value of $N$ would set our computational cost in
that it is the number of times that $S(i)$ must be queried. However, it
turns out that there are $N_{C}=85439$ cyclotomic cosets, and this is the
actual number of queries to $S(i)$.

Note that there are instances where $N\ll n$ and cyclotomic cosets are not
required. For example consider the binary $[13981,20]$ irreducible cyclic
code. Here $n=13981$ and $N=75$. Physically this corresponds to either the
fully anti-ferromagnetic or ferromagnetic Ising model over a connected graph
with 13981 edges and 13962 vertices (considering the dual code).

\section{Classical and Quantum Complexity of the Scheme}

\label{complexity}

Assuming one knew that a given graph was a member of $\mathrm{ICCC}%
_{\epsilon }$, then classically one could proceed as follows using a state
of the art algorithm ZETA for the computation of zeta functions of the
family of curves $C_{\alpha }:y^{q}-y=\alpha x^{N}$ \cite{Denef:04}. Here $N$
is as given in the relation $nN=q^{k}-1$ and the index $\alpha $ is in
one-one correspondence with the code words in the given cocycle code
(specifically $\alpha \in \mathbf{F}_{q^{k}}$). The connection between the
weights of words of an irreducible cyclic code and the number of rational
points on the curves $C_{\alpha }$ is well known, as is the connection
between the zeta functions of such curves and Gauss sums \cite{vlugt}. The
complexity of using ZETA to compute the $N=\alpha k^{s(k)}$ different
weights is $O(k^{6s(k)+3+\epsilon ^{\prime }}\left( \frac{q}{2}\right)
^{5+\epsilon ^{\prime }})$ \cite{Denef:04} and a tally of these weights will
take $O(k^{s(k)})$ operations ($\epsilon ^{\prime }$ is a small real number
-- unrelated to $\epsilon $ which parameterizes the class of graphs in
question). The overall complexity of finding the range of $S(i)$ will
therefore be 
\begin{equation}
\text{classical cost}=O(k^{6s(k)+3+\epsilon ^{\prime }}\left( \frac{q}{2}%
\right) ^{5+\epsilon ^{\prime }}),  \label{eq:cc}
\end{equation}%
assuming that we know that a given graph is a member of $\mathrm{ICCC}%
_{\epsilon }$. As far as we know this is the fastest classical algorithm for
the problem we have considered here.

For a quantum computer we do not need to assume that testing for membership
is efficient:\ we know that this can be done efficiently using the discrete
log algorithm \cite{Shor:97}. Above we showed that the overall complexity of
finding $Z$ is bounded by $O(k^{2\max [1,s(k)]}(\log q)^{2})$. This should
be contrasted with the best classical result available, (\ref{eq:cc}). For
example if we take $s=2$, (both classical and quantum methods are polynomial
when we take $s(k)$ to be a constant) we obtain an $O(k^{11})$ improvement
and an exponential speedup in $q$. One could imagine fixing a graph and
calculating the partition function for increasing values of $q$. In this
situation we have an exponential speedup over the best classical algorithm
available.

Note that there is a quantum algorithm for finding zeta functions of curves
which is exponentially faster in $q$ than the classical algorithm in \cite%
{Denef:04} (as is ours). This is given in \cite{Kedlaya:05}. The use of this
algorithm instead of the Gauss sum approximation algorithm is left for a
future publication.

On a final note, the classification $\mathrm{ICCC}_{\epsilon }$ we have
chosen is meant to highlight the boundary between $BQP$ and $P$ by fixing
the acceptable error in the Gauss sum phases. One could opt for a perhaps
more natural class of graphs by bounding the way that $1/\epsilon $ grows
instead. For example, one could restrict the class of graphs in such a way
that $1/\epsilon \sim q^{\frac{k}{2}-\theta _{n,k}+1}$ grows polynomially in 
$k$, in particular such that%
\begin{equation*}
\frac{1}{\epsilon }<k^{5s(k)+1}.
\end{equation*}
For this class of graphs one would also have a speedup in the quantum case.

\section{Detailed Summary}

\label{summary}

\begin{figure}
\epsfxsize=14cm
\centerline{\epsffile{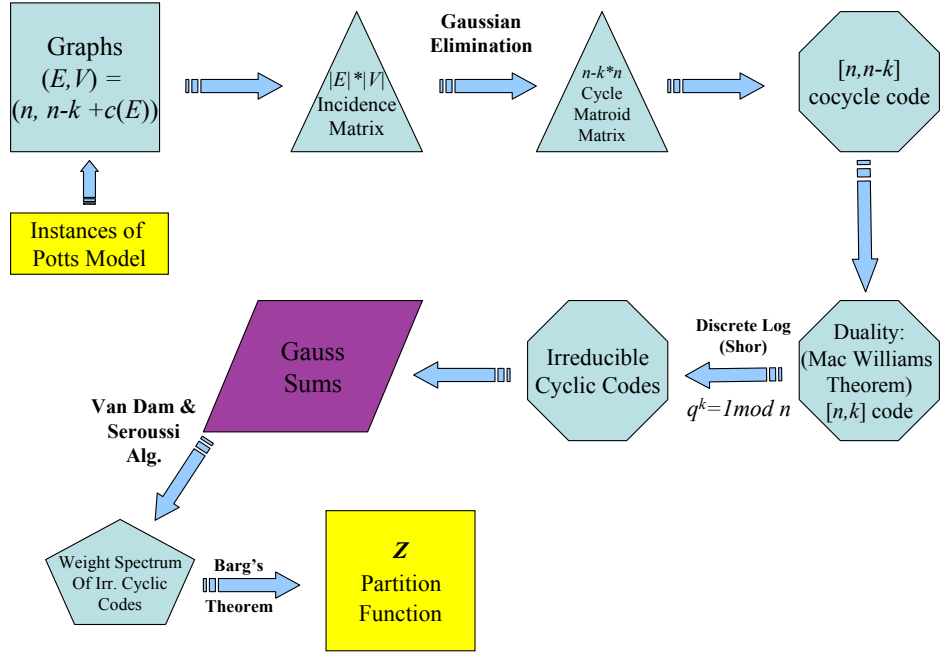}}
\caption{A diagrammatic overview of the algorithm. (Box shapes do not
  have a meaning.)}
\end{figure}

For convenience we recollect our definitions and provide a diagram of our
scheme. We are considering the $q$-state Potts model (fully ferromagnetic or
fully anti-ferromagnetic) over a graph $\Gamma =(E,V)$, with $q$ prime. This
includes the Ising model ($q=2\,$). Every graph $\Gamma $ has a cycle
matroid $M(\Gamma )$ associated with it and every cycle matroid has a $%
(|V|-c(\Gamma ))\times |E|$ matrix representation $G$ (the CMM), where $%
c(\Gamma )$ is the number of connected components of $\Gamma $. The columns
of $G$ encode the dependence structure of the graph and the row space of $G$
generates the cocycle code of length $|E|$ and dimension $|V|-c(\Gamma )$.
The length and dimension of the dual code are respectively $n=|E|$ and $%
k=|E|-|V|+c(\Gamma )$.

Following is a detailed synopsis of the algorithm for computing the
partition function.

\begin{enumerate}
\item Given a graph, efficiently determine if it belongs to $\mathrm{ICCC}%
_{\epsilon }$ (Definition \ref{iccc}). This step appears to be hard on a
classical computer in general, since it is equivalent to computing a
discrete log.

\item If the CMM $G=[I_{|V|-c(\Gamma )}|X]$ is the matrix representation
over $\mathbf{F}_{q^{k}}$ of the cycle matroid of $\Gamma $, $M(\Gamma )$,
then the row space of $H=[-X^{T}|I_{|E|-|V|+c(\Gamma )}]$ will be the code $%
C(\Gamma )$.

\item Let $N=O(k^{s})$ where $s$ is a constant integer that determines the
complexity of the algorithm. Take $C(\Gamma )$ as an irreducible cyclic code
of length $n=\frac{q^{k}-1}{N}$ and dimension $k$, i.e., we only consider
graphs $\Gamma $ where $C(\Gamma )$ is an irreducible $[n,k]$ cyclic code.

\item If we can evaluate the weight enumerator of $C(\Gamma )$ we will have
successfully approximated the Potts partition function over the
corresponding graph $\Gamma $. To do so:

\begin{enumerate}
\item Find the $q$-cyclotomic cosets of $\{0,1,\dots ,N-1\}$. This step
requires at most linear time in $N$.

\item Using the quantum algorithm for Gauss sums \cite{van} we are be able
to estimate the weights of the words. The error in the Gauss sum algorithm
can be high in this setting, and therefore we have to restrict the class of
graphs further in order to obtain exact evaluations. Use the Gauss sum
algorithm to return the phases $\gamma _{1},\dots ,\gamma _{d-1}$ [Eq. (\ref%
{gauss})] and then input these values into the function $S(i)$ [Eq. (\ref%
{S(i)})]. According to the McEliece Theorem (Th.~\ref{mceliece}) we have to make $%
d-1$ (where $d=\gcd (N,\frac{q^{k}-1}{q-1})$) calls to the quantum oracle
and we can use these evaluations for each representative $i$ of the $q$%
-cyclotomic cosets of $\{0,1,\dots ,N-1\}$. This step has time complexity $%
O(dk^{2}(\log q)^{2})$.

\item Let $b_{1},b_{2},\dots ,b_{N_{C}}$ be the coset representatives from the $N_{C}$ cosets. Now each coset has cardinality $%
v_{i}$, i.e., $b_{i}$ belongs to coset $i$ which has $v_{i}$ elements. We
evaluate $\omega _{i}=S(b_{i})$ for each $b_{i}$, remembering that each $%
\omega _{i}$ occurs $v_{i}$ times. We end up with a list $(\omega
_{1},\omega _{2},\dots ,\omega _{N_{C}})$ as well as a list $%
(v_{1},v_{2},\dots ,v_{N_{C}})$ of multiplicities.

\item Now perform a tally of repeats of the $\omega _{i}$ for each $i\in
\{1,...,N_{C}\}$. This returns a set of indices $\Lambda _{i}\equiv
\{j_{i}\}\subseteq \{1,...,N_{C}\}$. We add the corresponding $v_{j_{i}}$
which yields $a_{i}=\sum_{j\in \Lambda _{i}}v_{j}$, the number of words of
weight $\omega _{i}$ up to cyclic permutations. To account for cyclic
permutations due to the fact that we are working over cyclic codes, we have $%
A_{i}=na_{i}$, which is the desired weight spectrum.
\end{enumerate}

\item Now that we have determined the weight spectrum $A_{i}$ in time $%
O(k^{2s}(\log q)^{2})$ we have the coefficients for $A(1,y)$ and so via the
MacWilliams identity (\ref{mac}) we finally obtain the partition function (%
\ref{Zfinal}).
\end{enumerate}

\section{Conclusions and Future Directions}

In this work we have given a quantum algorithm for the exact evaluation of the
fully ferromagnetic or anti-ferromagnetic Potts partition function $Z$ under
the restriction to certain sparse graphs (with logarithmically more
edges than vertices). The methods we used exploit the
connection between coding theory and statistical physics. The motivation for
this work is an ongoing effort to identify instances of classical
statistical mechanics for which quantum computers will have an advantage
over classical machines.

The approach we described involves using the link between classical coding
theory and the Potts model via the weight enumerator polynomial $A$. One
should note that $A$ is another instance of the Tutte polynomial and so this
connection is not surprising. The weight enumerator encodes information
about all the different Hamming weights of the code words in a linear code
and the weight of a code word can be given by a formula involving a sum of
Gauss sums when dealing with a specific type of linear code. Since there
exists an efficient algorithm to approximate Gauss sums via quantum
computation \cite{van} we were able to efficiently calculate the weights
of code words for certain codes. Much of this paper dealt with the necessary
restrictions that one must impose in order to achieve this last step. For
example, once an error $\epsilon $ in the Gauss sum algorithm is
accepted, we demonstrated that there is a family of graphs for which one can
find the exact partition function, and therefore the error does not scale
within this family. Given a graph $\Gamma $, one can map the graph to a
corresponding linear code via the incidence structure of $\Gamma $. The
Potts partition function of $\Gamma $ (with either fully ferromagnetic or
anti-ferromagnetic interactions) is given by some easily computed function
times the weight enumerator of the corresponding code. Due to the symmetries
inherent in the mathematical structure of linear codes we were able to
provide an efficient method to exactly determine $Z$ for a class of graphs ($%
\mathrm{ICCC}_{\epsilon }$) which has a well defined correspondence to a
subset of linear codes.

In \cite{vyalyi} it was shown that the exact evaluation of weight
enumerators for binary linear codes is hard for the polynomial
hierarchy. As our approach involved the exact evaluation of weight
enumerators, it is not surprising that we had to make restrictions on
the class of graphs so as to make our scheme efficient. The vantage that coding theory gives to this particular
problem, however, allows one to utilize the fact that certain graphs have
properties that a quantum computer can take advantage of to provide a speed
up.

Notice that the related results in \cite{Aharonov:06,Aharonov:07} concern 
\emph{additive approximations}; the methods used in this paper can be
extended to a wider class of graphs if one relaxes the requirement of exact
evaluation and instead similarly considers additive approximations of $Z$.
An open question is what instances of the Potts partition function are
amenable to an fpras (fully polynomial random approximation scheme). The
methods used in \cite{Aharonov:06,Aharonov:07} have proven to be quite
powerful. There is hope to extend some of these methods to non-planar
graphs. One idea is to extend the algorithm in \cite{Aharonov:06} to the
Jones polynomial for virtual knots and then use some correspondence between
the virtual knots and non-planar graphs. Another approach may involve seeing
things in a new light. Note that the Jones polynomial is the Euler
characteristic of a certain chain complex \cite{Khovanov:00}. One can
explore how effective quantum computers will be at approximating Euler
characteristics in general. Perhaps there is a way of exploiting this in
order to obtain knowledge about the Potts partition function.

One may also consider strengthening the results given here by exploiting
theorems about the minimal distance of cyclic codes. For example, there are
theorems that guarantee a lower bound for the weight between any two words.
By enforcing that the generator polynomial of the code be of a certain form,
one would be guaranteed a certain distance between words and therefore the
error in the Gauss sum approximation will be of little consequence for
certain graphs \cite{Lint:book}. As already mentioned in the Introduction,
another potentially promising approach is to consider the scheme we have
presented here but to replace the Gauss sum algorithm with the quantum
algorithm for obtaining Zeta functions \cite{Kedlaya:05}. Work has to be
done on understanding the exact cost of this algorithm when one is
restricted to curves that are pertinent for the evaluation of the Potts
model.

Corollary~\ref{cor} deals with the combination of graphs via a
direct sum of codes gives one a way of \textquotedblleft
tiling\textquotedblright\ graphs for which one knows the partition function.
This gives a quick way of obtaining the partition function of certain graphs
that are made of many repeats of a simpler graph. There are other ways of
combining codes that may allow one to study the partition function of new
graphs, for example the concatenation or direct product of two codes \cite%
{Lint:book}.

The coding theoretic approach does give us a way of evaluating the partition
function of instances of the Potts model at arbitrary temperatures but
precisely the kinds of graphs which are involved is a question for future
research. Indeed, the identification of the physical instances represented
by the graphs for which our algorithm is efficient will shed light on the
question that motivated this work in the first place \cite{Lidar:QWGT}:
what is the quantum computational complexity of classical statistical
mechanics?

\acknowledgments
J.G. would like to thank Marko Moisio, Ravi Minhas, and Frank van
Bussel for helpful discussions. D.A.L. gratefully acknowledges support under ARO
grant W911NF-05-1-0440.


\appendix

\section{Matroids}
\label{appA}

\begin{mydefinition}
A matroid $M$ on a set $E$ is the pair $(E,I)$ where $I$ is a collection
of subsets of $E$ with the following properties:

\begin{enumerate}
\item The empty set is in $I$.

\item Hereditary Property: If $A\in I$ and $B\subset A$, then $B\in I$.

\item Exchange Property: If $A$ and $B$ are in $I$ and $A$ has more elements
than $B$, then $\exists a\in A$ such that $a\notin B$ but $B\cup \{a\}\in I$.
\end{enumerate}
\end{mydefinition}

The collection of sets in $I$ are called the independent sets and $E$ is
referred to as the ground set.

\begin{mydefinition}
A cycle matroid of a graph $\Gamma $ is the set of all edges of $\Gamma $ as
the ground set $E$ together with $I$ as the subsets of $E$ which do not
contain a cycle. So the independent sets are collections of edges which do
not have cycles.
\end{mydefinition}

Recall that in graph theory one refers to such an edge set (the above
independent set) as a forest.

In matroid theory a matrix representation is a matrix whose column vectors
have the same dependence relations as the matroid it is representing. More clearly, the column vectors represent the matroid elements and the
usual notion of linear dependency determines the dependent sets and
therefore the independent sets as well. Thus, the matrix can be said to
generate the matroid.

As an example, imagine the triangle graph of three nodes with three edges $A$%
,$B$, and $C$. The cycle matroid consists of each of the edges individually
and any collection of two edges. All three edges form a cycle so it cannot
be included. We require our matrix representation to encode this
independence structure of the edges. One may work over any field here
because we are only concerned with graphic matroids, i.e., matroids which
can be represented as a cycle matroid of some graph. (Graphic matroids are
representable over any field \cite{Welsh:matroid}.) Now, if we think of
column 1,2 and 3 as edges $A$,$B$ and $C$ respectively we can take the
following matrix as a representation in $\mathbf{F}_{2}$:

\begin{equation*}
\left( 
\begin{array}{ccc}
1 & 0 & 1 \\ 
0 & 1 & 1%
\end{array}
\right)
\end{equation*}

Since addition is mod $2$ here, a cycle is any collection of columns
that sum to the $0$-vector. We can take all collections where this does
not happen and these collections will form $I$. In this way, this matrix is
a representation of the cycle matroid for the triangle graph. In matroid
theory one has the familiar notion of a base.

\begin{mydefinition}
A base of a matroid $M=(E,I)$ is a maximal independent subset of $E$.
\end{mydefinition}

It is not a coincidence that the left part of the matrix is the $2\times 2$
identity matrix. In general one can form a representation (known as the
standard matrix representation) where one begins with an identity matrix
which is $r\times r$ where $r$ is the size of the base of $M$ and append to
it columns that capture the dependence structure of the matroid in question.
In this way, the columns of the identity matrix represent the chosen basis of 
$M$. So $M$ is isomorphic to the matroid induced on the columns of the
matrix by linear dependence. A more precise explanation can be found in \cite%
{Welsh:matroid}. What is important for us is that such a matrix
representation is possible.

\section{Algebraic approach to (Irreducible) Cyclic Codes}

\label{appB}

\subsection{Irreducible cyclic codes as minimal ideals}

Let us recall some definitions from algebra. Take $q$ to be prime or a power of a prime.

\begin{mydefinition}
A ring is a set $R$ which is an abelian group $(R,+)$ with $0$ as the
identity, together with $(R,\times )$, which has an identity element with
respect to $\times $ where $\times $ is associative.
\end{mydefinition}

\begin{mydefinition}
An ideal $I$ is a subset of a ring $R$ which is itself an additive subgroup
of $(R,+)$ and has the property that when $x\in R$ and $a\in I$ then $xa$
and $ax$ are also in $I$.
\end{mydefinition}

\begin{mydefinition}
A principle ideal is an ideal where every element is of the form $ar$ where $%
r\in R$.
\end{mydefinition}

Thus, a principle ideal is generated by the one element $a$ and a principal
ideal ring is a ring in which every ideal is principle.

There is an important isomorphism between powers of finite fields $\mathbf{F}
_{q}^{n}$ and a certain ring of polynomials. Recall that the multiples of $%
x^{n}-1$ form a principal ideal in the polynomial ring $\mathbf{F}_{q}[x]$.

Therefore the residue class ring $\mathbf{F}_{q}[x]/(x^{n}-1)$ is isomorphic
to $\mathbf{F}_{q}^{n}$ since it consists of the polynomials 
\begin{equation*}
\{a_0 + a_1x + \cdots + a_{n-1} x^{n-1} | a_i \in \mathbf{F}_q, 0 \le i <n
\}.
\end{equation*}

Therefore, taking multiplication modulo $x^n -1$ we can make the following
identification: 
\begin{equation}
(a_0,a_1,\dots ,a_{n-1}) \in \mathbf{F}_q^n \longleftrightarrow a_0 + a_1x +
\cdots + a_{n-1} x^{n-1} \in \mathbf{F}_q[x]/(x^n-1) .  \label{corr}
\end{equation}

This implies the following theorem.

\begin{mytheorem}
A linear code $C$ in $\mathbf{F}_{q}^{n}$ is cylic $\iff $ $C$ is an ideal
in $\mathbf{F}_{q}[x]/(x^{n}-1)$.\cite{Lint:book}
\end{mytheorem}

\begin{proof}
In one direction this is easy since if $C$ is an ideal in $\mathbf{F}
_{q}[x]/(x^{n}-1)$ and $c(x)=a_{0}+a_{1}x+\cdots +a_{n-1}x^{n-1}$ is a
codeword, then by definition $xc(x)\in C$ as well and so $%
(a_{n-1},a_{0},a_{1},\dots ,a_{n-2})\in C$. In the other direction, one just
has to note that since $C$ is cyclic, $xc(x)$ is in $C$ for every $c(x)\in C$
which means that $x^{k}c(x)$ is in $C$ for every $k$. But $C$ is linear by
assumption so if $h(x)$ is any polynomial then $h(x)c(x)$ is in $C$ and thus 
$C$ is an ideal.
\end{proof}

Note that $\mathbf{F}_{q}[x]/(x^{n}-1)$ is a principal ideal ring and
therefore the elements of every cyclic code $C$ are just multiples of $g(x)$, the monic polynomial of lowest degree in $C$; $g(x)$ is called the
generator polynomial of $C$. Because of the correspondence (\ref{corr})
above we know that given $g(x)=g_{0}+g_{1}x+\cdots g_{n-k}x^{n-k}$ [$g(x)$
divides $x^{n}-1$ since otherwise $g(x)$ could not be the monic polynomial
of lowest degree in $C$], we have the vector $(g_{0},g_{1},\dots ,g_{n-k})$.
We then can write the $k\times n$ generator matrix of the code as 
\begin{equation*}
\left( 
\begin{array}{cccccccc}
g_{0} & g_{1} & \cdots & g_{n-k} & 0 & 0 & \cdots & 0 \\ 
0 & g_{0} & \cdots & g_{n-k-1} & g_{n-k} & 0 & \cdots & 0 \\ 
0 & 0 & \cdots &  &  &  & \cdots & 0 \\ 
0 & 0 & \cdots &  & g_{0} & g_{1} & \cdots & g_{n-k}%
\end{array}
\right) .
\end{equation*}

In this way, the row space generates $C$. If we can
write $x^{n}-1=w_{1}(x)w_{2}(x)\cdots w_{t}(x)$ as the decomposition of $%
x^{n}-1$ into irreducible factors, then the code generated by $\frac{x^{n}-1%
}{w_{i}(x)}$ is called an \emph{irreducible} cyclic code. In algebraic terms
what this means is that the code $C$ is a \emph{minimal ideal} of $\mathbf{F}%
_{q^{k}}[x]/(x^{n}-1)$, i.e., $C$ contains no subspace (other than $0$)
which is closed under the cyclic shift operator \cite{Moiso:97}. The reason
we are interested in irreducible cyclic codes is that there is an
established connection between the weights of the code words and Gauss sums.

We now turn to the representation of irreducible cyclic codes, specifically
1) the form that the generator matrix can take, 2) a description of the
codewords in terms of the trace function. Issue 1) relates back to the
matrix representation of the cycle matroid of graphs and issue 2) will allow
us to make the connection to Gauss sums.

\subsection{Generator matrix of a cyclic code and the cycle matroid matrix}

There is an alternative (but equivalent) way of constructing the generator
matrix of a cyclic code which will immediately show its usefulness in its
relationship with the cycle matroid matrix representation. Let $C$ be an $%
[n,k]$ cyclic code and let $g(x)$ be the generator polynomial. Now, divide $%
x^{n-k+i}$ by $g(x)$ for $0\leq i\leq k-1$. We have 
\begin{equation*}
x^{n-k+i}=q_{i}(x)g(x)+r_{i}(x)
\end{equation*}%
where $\deg r_{i}(x)$ $<\deg g(x)=n-k$ or $r_{i}(x)=0$. What this means is
that we have a set of linearly independent code words. Namely, we have the $%
k $ code words given by 
\begin{equation*}
x^{n-k+i}-r_{i}(x)=q_{i}(x)g(x)
\end{equation*}%
in $C$. More explicitly, take the remainder polynomials $r_{i}(x)$ after
applying the division algorithm and using the correspondence (\ref{corr})
above, form the $k\times (n-k)$ matrix $R$ and append the $k\times k$
identity matrix to it. The rows of $R$ are the coefficients of the $r_{i}(x)$
and one then has the $k\times n$ generator matrix $[I_{k}|R]$. This is
precisely the form of the matrix representation for matroids discussed
above. Thus, we have a correspondence between the generator matrix for
an irreducible cyclic code and the matrix representation for the cycle
matroid of a graph.

\begin{myproposition}
\label{prop:N}In an $[n,k]$ irreducible cyclic code there are at most $N$
words of different non-zero weight where $N=(q^{k}-1)/n$.
\end{myproposition}

\begin{proof}
For any irreducible cyclic code we have the relation $q^k-1 = nN$ over the
field $\mathbf{F}_q$. The length of each word is $n$ and any cyclic
permutation of a word preserves the Hamming weight. Therefore, for each word
there are $n-1$ other words of equal weight. As there are $q^{k}-1$ words of
non-zero weight, if we assume that every word that does not arise from the
cyclic permutation of another word is of a different weight, then there are $%
(q^{k}-1)/n$ words of different weight. Being however that there is the
possibility of repeats in weight among words which are not cyclic
permutations of each other, there are at most $N$ different weights.
\end{proof}

\section{Gauss Sums and a Quantum Algorithm for the estimation of Gauss Sums}

\label{appC}

Gauss sums are sums of products of group characters.

\subsection{Characters}

\label{app:Characters}

A character of a finite group $(G,\ast )$ is a homomorphism $\Phi $ from $G$
to the group of the non-zero complex numbers $\mathbf{C}$.

We are interested in two types of characters, namely the multiplicative and
additive characters. Let $\mathbf{F\equiv F}_{q^{k}}$ (where $k$ is a
positive integer) be a finite field as defined previously, and let $\mathbf{F%
}^{\ast }$ be the multiplicative group of $\mathbf{F}$. Let $g$ be a
primitive element of $\mathbf{F}$ (i.e., $g$ generates $\mathbf{F}$). Let 
\begin{equation*}
\zeta _{q}=e^{2\pi i/q}
\end{equation*}%
denote the $q$th root of unity. Let $x=g^{k}\in \mathbf{F}^{\ast }$. A
multiplicative character $\chi _{j}(x)$ is a mapping from the set of powers $%
\{m\}$ in $x=g^{m}$ to powers of roots of unity. Specifically, the group of
multiplicative characters $\chi =\{\chi _{j}\}_{j}$ consists of the elements 
\begin{equation*}
\chi _{j}(x)=\chi _{j}(g^{m})=\zeta _{q^{k}-1}^{jm},\hspace{0.25in}m=0,\dots
,q-2\in \mathbf{F}_{q};\emph{\quad }j=0,\dots ,q^{k}-2\in \mathbf{F}.
\end{equation*}%
Let $a\in \mathbf{F}$. An additive character $e_{j}(a)$ is a mapping from $%
\mathbf{F}$ to powers of roots of unity via the trace function.
Specifically, the group of additive characters $e=\{e_{\beta}\}_{\beta}$ consists of
the elements 
\begin{equation*}
e_{\beta}(a)=\zeta _{q}^{\mathrm{Tr}(\beta a)}\hspace{0.25in}\forall a,\beta=0,\dots
,q^{k}-1\in \mathbf{F}
\end{equation*}%
where the trace is defined in Eq.~(\ref{eq:Tr}).

\subsection{Discrete Log}

For every non-zero $x\in \mathbf{F}^{\ast }$ the discrete logarithm with
respect to a primitive element $g\in \mathbf{F}$ is given by 
\begin{equation*}
\log _{g}(x)=\log _{g}(g^{m})=m\, \mathrm{mod}\, (q^{k}-1).
\end{equation*}%
This means that every multiplicative character can be written 
\begin{equation}
\chi _{j}(x)=\chi _{j}(g^{m})=\zeta _{q^{k}-1}^{j\log _{g}(x)}  \label{chi}
\end{equation}%
for $x\neq 0$ and $\chi (0)=0$.

\subsection{Gauss Sums}

Let $e_{\beta}$ and $\chi _{j}$ be an additive and multiplicative character
respectively. Then the Gauss Sum $G(\chi _{j},e_{\beta})$ is defined as: 
\begin{equation}
G(\chi _{j},e_{\beta})=\sum_{x\in \mathbf{F}^{\ast }}\chi _{j}(x)e_{\beta}(x).
\label{sum}
\end{equation}
Gauss sums are used extensively in number theory, e.g., in the study of quadratic
residues or Dirichlet L-functions.

To compute a Gauss sum we need to specify the field $\mathbf{F}$ and the
indices $\beta \in \mathbf{F}$ and $j\in \mathbf{F}$ of the additive and
multiplicative characters respectively. Thus the input size to a Gauss sum
computation is $O(k \log q)$ bits. We can now define the Gauss sum over $%
\mathbf{F}$ as 
\begin{equation*}
G_{\mathbf{F}}(\chi _{j},\beta)=\sum_{x\in \mathbf{F}^{\ast }}\chi _{j}(x)\zeta
_{q}^{\mathrm{Tr}(\beta x)}.
\end{equation*}

It is well known that if $\chi_j \neq 1$ then \cite{Berndt:book}: 
\begin{equation}
G_{\mathbf{F}}(\chi _{j},\beta)=\sqrt{q^{r}}e^{i\gamma },
\end{equation}
where $\gamma =\gamma _{\mathbf{F}}(\chi _{j},\beta)$. This means that all we
need to do is approximate the angle $\gamma \, \mathrm{mod}\, (2\pi )$ in order to
approximate the Gauss sum. This is precisely the Gauss sum approximation
problem for finite fields.

\subsection{Quantum Algorithm for Gauss Sums}

Van Dam and Seroussi devised an efficient quantum algorithm to estimate
Gauss Sums \cite{van}. The following is an outline of the essentials of the
proof; we refer the reader to \cite{van} for a complete description as well
as a discussion of the complexity of estimating Gauss sums.

\begin{mytheorem}
\textbf{\{Quantum Amplitude Amplification\}} Let $f:S\mapsto \{0,1\}$ be a
function for which we know the total weight $\lVert f\rVert _{l_{1}}$ but
not those values $x\in S$ for which $f(x)=1$. Then the corresponding state 
\begin{equation*}
|f\rangle =\frac{1}{\lVert f\rVert _{l_{2}}}\sum_{x\in S}f(x)|x\rangle
\end{equation*}%
can be efficiently and exactly prepared on a quantum computer where we have
to make a number of queries to $f$ of the order $O\left( \sqrt{\frac{\lvert
S\rvert }{\lVert f\rVert _{l_{1}}}}\right) $.
\end{mytheorem}

This is an essential ingredient in Grover's quantum search algorithm. For a
proof and details see \cite{Grover:96}. It follows from Eq.~(\ref{chi}) and
Shor's discrete log algorithm \cite{Shor:97} that given $g$, $q^{k}$ and $j$%
, we can efficiently create the state $|\chi _{j}\rangle $. The following
lemma is essential in this regard. First note that for any set $S$ we define 
\begin{equation*}
|S\rangle \equiv \frac{1}{\sqrt{\lvert S\rvert }}\sum_{x\in S}|x\rangle .
\end{equation*}

\begin{mylemma}
For a finite field $\mathbf{F}_{q^{k}}$ and the triplet $(q^{k},g,r)$ (the
specification of a multiplicative character $\chi _{r}$), the state 
\begin{equation*}
|\chi _{r}\rangle =\frac{1}{\sqrt{q^{k}-1}}\sum_{x\in \mathbf{F}%
_{q^{k}}}\chi _{r}(x)|x\rangle
\end{equation*}%
and its Fourier transform $|\hat{\chi _{r}}\rangle $ can be created in $%
\mathrm{polylog}(q^{k})$ time steps on a quantum computer.
\end{mylemma}

\begin{proof}
We first create the state 
\begin{equation*}
|\mathbf{F}_{q^{k}}^{\ast }\rangle |\hat{1}\rangle =\frac{1}{\sqrt{%
q^{k}(q^{k}-1)}}\sum_{x\in \mathbf{F}_{q^{k}}^{\ast }}|x\rangle
\sum_{j=0}^{q^{k}-2}\zeta _{q^{k}-1}^{j}|j\rangle
\end{equation*}%
by using Grover's amplitude amplification on $\mathbf{F}_{q^{k}}$ and the
Fourier transform. Next, in superposition over all $x\in \mathbf{F}%
_{q^{k}}^{\ast }$, we calculate $\log _{g}(x)$ and subtract $r\log _{g}(x)$.%
\begin{eqnarray}
|\mathbf{F}_{q^{k}}^{\ast }\rangle |\hat{1}\rangle &\longrightarrow &\frac{1%
}{\sqrt{q^{k}(q^{k}-1)}}\sum_{x\in \mathbf{F}_{q^{k}}^{\ast }}|x\rangle
\sum_{j=0}^{q^{k}-2}\zeta _{q^{k}-1}^{j}|j-r\log _{g}(x)\rangle \\
&=&\frac{1}{\sqrt{q^{k}(q^{k}-1)}}\sum_{x\in \mathbf{F}_{q^{k}}^{\ast
}}|x\rangle \sum_{j=0}^{q^{k}-2}\zeta _{q^{k}-1}^{j}\zeta _{q^{k}-1}^{r\log
_{g}(x)}|k\rangle \\
&=&\frac{1}{\sqrt{q^{k}-1}}\sum_{x\in \mathbf{F}_{q^{k}}^{\ast }}\zeta
_{q^{k}-1}^{r\log _{g}(x)}|x\rangle |\hat{1}\rangle \\
&=&|\chi _{r}\rangle |\hat{1}\rangle
\end{eqnarray}%
To get $|\hat{\chi _{r}}\rangle $ we just need to apply the Fourier
transform.
\end{proof}

The technique used in the above proof is known as the \emph{phase kickback
trick}. Now we are ready for the following.

\begin{mytheorem}
\textbf{Algorithm for approximating Gauss Sums.} Consider $\mathbf{F}%
_{q^{k}} $, a nontrivial multiplicative character $\chi _{r}$ and $\beta \in 
\mathbf{F}_{q^{k}}^{\ast }$. If we apply the quantum Fourier transform over
this field to $|\chi _{r}\rangle $, followed by a phase change 
\begin{equation}
|y\rangle \longrightarrow \chi _{r}^{2}(y)|y\rangle  \label{phase}
\end{equation}%
then we generate an overall phase change given by 
\begin{equation*}
|\chi _{r}\rangle =\frac{1}{\sqrt{q^{k}-1}}\sum_{x\in \mathbf{F}%
_{q^{k}}}\chi _{r}(x)|x\rangle \longrightarrow \frac{G_{\mathbf{F}%
_{q^{k}}}(\chi _{r},\beta)}{\sqrt{q^{k}}}|\chi _{r}\rangle .
\end{equation*}
\end{mytheorem}

\begin{proof}
After a Fourier transform we have 
\begin{eqnarray*}
|\hat{\chi _{r}}\rangle &=&\frac{1}{\sqrt{q^{k}(q^{k}-1)}}\sum_{y\in \mathbf{%
F}_{q^{k}}^{\ast }}\left( \sum_{x\in \mathbf{F}_{q^{k}}}\chi _{r}(x)\zeta
_{q}^{\mathrm{Tr}(\beta xy)}\right) |y\rangle \\
&=&\frac{1}{\sqrt{q^{k}(q^{k}-1)}}\sum_{y\in \mathbf{F}_{q^{k}}^{\ast }}{G_{%
\mathbf{F}_{q^{k}}}(\chi _{r},\beta y)|y\rangle } \\
&=&\frac{1}{\sqrt{q^{k}(q^{k}-1)}}\sum_{y\in \mathbf{F}_{q^{k}}^{\ast }}\chi
_{r}(y^{-1})G_{\mathbf{F}_{q^{k}}}(\chi _{r},\beta)|y\rangle .
\end{eqnarray*}%
Then 
\begin{equation*}
|\chi _{r}\rangle =\frac{G_{\mathbf{F}_{q^{k}}}(\chi _{r},\beta)%
}{\sqrt{q^{k}(q^{k}-1)}}\sum_{y\in \mathbf{F}_{q^{k}}^{\ast }}\chi
_{r}(y^{-1})|y\rangle .
\end{equation*}%
Now we know that we can efficiently (and exactly) create the phase change
given by (\ref{phase}). Doing so gives us 
\begin{equation*}
|\hat{\chi}\rangle \longrightarrow \frac{G_{\mathbf{F}_{q^{k}}}(\chi _{r},\beta)%
}{\sqrt{q^{k}(q^{k}-1)}}\sum_{y\in \mathbf{F}_{q^{k}}^{\ast }}\chi
_{r}(y^{-1})\chi _{r}^{2}(y)|y\rangle =\frac{G_{\mathbf{F}_{q^{k}}}(\chi
_{r},\beta)}{\sqrt{q^{k}}}|\chi _{r}\rangle
\end{equation*}%
since $|\chi _{r}\rangle =\frac{1}{\sqrt{q^{k}-1}}\sum_{y\in \mathbf{F}%
_{q^{k}}^{\ast }}\chi _{r}(y)|y\rangle $ and $\chi _{r}(y^{-1})\chi
_{r}(y)=1 $. Thus, the coefficient of $|\chi _{r}\rangle $ is just $%
e^{i\gamma }$. It is well known that one can efficiently estimate the phase
of such a function to within an expected error of $O(1/n)$ where $n$ is the
number of copies of $e^{i\gamma }|\chi _{r}\rangle $ we sample. Therefore we
arrive at an estimate of $\gamma $ and hence of the Gauss sum in question.
\end{proof}

This gives way to the following theorem about the time complexity of the
algorithm and is the culmination of the first part of the paper \cite{van}.

\begin{mytheorem}
For any $\epsilon >0$, there is a quantum algorithm that estimates the phase 
$\gamma $ in $G_{\mathbf{F}_{q^{k}}}(\chi _{r},\beta)=\sqrt{q^{k}}e^{i\gamma }$,
with expected error $E(\lvert \gamma -\tilde{\gamma}\rvert )<\epsilon $. The
time complexity of this algorithm is bounded by $O(\frac{1}{\epsilon }\cdot 
\mathrm{polylog}(q^{k}))$ \cite{van}.
\end{mytheorem}

Note that the \textquotedblleft poly\textquotedblright\ in polylog refers to
a quadratic polynomial.

\end{document}